\documentclass{AGUJournal}

\journalname{Journal of Advances in Modeling Earth Systems}

\usepackage{amsmath}
\usepackage{epsfig}
\usepackage{url}
\usepackage{colordvi}
\usepackage{color}
\usepackage{txfonts}
\usepackage{graphicx}
\usepackage{natbib}
\usepackage{bm}
\usepackage{xspace,url}
\usepackage[toc,page]{appendix}
\usepackage{hyperref}
\graphicspath{{./fig/}}

\begin{document}

\title{Eulerian and modified Lagrangian approaches to multi-dimensional
condensation and collection}

\authors{Xiang-Yu Li\affil{1,2,3,4}, A.\ Brandenburg\affil{2,3,5,6}, N.\ E.\ L.\ Haugen\affil{7,8}, and G.\ Svensson\affil{1,3,9}}

\affiliation{1}{Department of Meteorology, and Bolin Centre for Climate Research, Stockholm University, Stockholm, Sweden}
\affiliation{2}{Nordita, KTH Royal Institute of Technology and Stockholm University, 10691 Stockholm, Sweden}
\affiliation{3}{Swedish e-Science Research Centre, www.e-science.se, Stockholm, Sweden}
\affiliation{4}{Laboratory for Atmospheric and Space Physics,
University of Colorado, Boulder, CO 80303, USA}
\affiliation{5}{JILA and Department of Astrophysical and Planetary Sciences
University of Colorado, Boulder, CO 80303, USA}
\affiliation{6}{Department of Astronomy, Stockholm University, SE-10691 Stockholm, Sweden}
\affiliation{7}{SINTEF Energy Research, 7465 Trondheim, Norway}
\affiliation{8}{Department of Energy and Process Engineering, NTNU, 7491 Trondheim, Norway}
\affiliation{9}{Global \& Climate Dynamics, National Center for Atmospheric Research, Boulder, CO 80305, USA}

\correspondingauthor{Xiang-Yu Li}{xiang.yu.li@su.se,~ $ $Revision: 1.715 $ $}

\begin{keypoints}
\item Eulerian Smoluchowski and Lagrangian superdroplet/superparticle
approaches to cloud droplet growth through condensation and collection
are compared using DNS techniques
\item Size spectra agree well for both approaches, especially in case
of turbulence
\item
The Lagrangian scheme with symmetric collection is found to be
optimal and computationally most efficient
\end{keypoints}

\begin{abstract}
Turbulence is argued to play a crucial role in cloud droplet growth.
The combined problem of turbulence and cloud droplet growth is numerically
challenging.
Here, an Eulerian scheme based on the Smoluchowski equation is compared with
two Lagrangian superparticle (or superdroplet) schemes in the presence
of condensation and collection.
The growth processes are studied either separately or in combination
using either two-dimensional turbulence, a steady flow, or just
gravitational acceleration without gas flow.
Good agreement between the different schemes for the time evolution of the
size spectra is observed in the presence of gravity or turbulence.
Higher moments
of the size spectra are found to be a useful tool to characterize the
growth of the largest drops through collection. Remarkably, the tails of the size
spectra are reasonably well described by a gamma distribution in cases
with gravity or turbulence.
The Lagrangian schemes are generally found to be superior
over the Eulerian one in terms of computational performance.
However, it is shown that the use of interpolation schemes
such as the cloud-in-cell algorithm is detrimental in
connection with superparticle or superdroplet approaches.
Furthermore, the use of symmetric over asymmetric collection schemes
is shown to reduce the amount of scatter in the results.
\end{abstract}

\section{Introduction}

In the context of raindrop formation,
it is generally accepted that turbulence plays a crucial role in
bridging the size gap between efficient condensational growth of small
particles (radii below $10\,\mu{\rm m}$) and efficient collectional
growth due to gravity of larger ones (radii around $100\,\mu{\rm m}$ and above)
\citep{Shaw03,Grabowski_2013,Khain2007}.
Improving the understanding of this important problem in meteorology
\citep{Berry74,Pinsky97,Falk02,Naumann16}
might also shed light on how to bridge the even more severe size gap
in the astrophysical context of planetesimal formation \citep{Joh07,Joh12}.
To address these questions numerically, one has to combine direct numerical
simulations (DNS) of turbulent gas motions with those of particles.
The particles are cloud droplets in the meteorological context and dust grains
in astrophysics.
A possible approach to treat collection is to solve the Smoluchowski
equation (also known as the stochastic collection equation in the
meteorological context)
\citep{OT73,SS02,Ray_2016}, which couples the spatio-temporal evolution
equations of the particle distribution function for different particle sizes.
The particle motion can be treated using a fluid description for each
particle size.
Thus, not only does one have to solve the Smoluchowski equation at each
meshpoint, but, because heavier particles have finite momenta and speeds
that are different from those of the gas, one has to solve corresponding
momentum equations for each mass species.
In the meteorological context, it is also referred to as
a binned spectral method, although in that case the momentum
equations for the particle bins are normally ignored \citep{Xue08}.
An Eulerian approach is technically more straightforward
than a Lagrangian one, but it becomes computationally demanding
as the size range of cloud droplets is large.

The Eulerian approach also has conceptual difficulties
if the collection probability depends on the mutual velocity difference.
This is due to the fact that particles of the same size are described by the
same momentum equation and have therefore the same velocity at a given
position in space, so the velocity difference vanishes.
This means that particles of the same size are not allowed to collide.
This is not a problem for freely falling particles of the same size,
which would
have the same terminal velocity and are not expected to collide.
This would however be an unrealistic restriction when particles
are subject to acceleration by turbulence.
More importantly, as was emphasized in the recent review of
\cite{Khain15}, the Smoluchowski equation is a mean-field equation and
cannot capture the random properties of the collections if the collision
kernel is prescribed a priori. Nevertheless, most numerical
cloud microphysical approaches are based on the Smoluchowski equation,
which therefore raises questions regarding the accuracy of the basic
equations \cite{Khain15}.
Thus, new approaches based on inherently different equations are required
to model the cloud microphysical processes.

An alternative approach is the Lagrangian one, where one solves for the
motion of individual particles and treats collections explicitly.
In atmospheric clouds, the number density of micrometer-sized cloud droplets
is of the order of $10^8\,{\rm m}^{-3}$, so in a volume of $1\,{\rm m}^3$, one has
100 million particles, which is the typical size that can be
treated on modern supercomputers.
A domain of this size is also about the largest that is possible in direct
numerical simulations (DNS) of atmospheric turbulence;
the Reynolds number based on the length scale $\ell=1\,{\rm m}$
and the corresponding velocity scale $u_\ell\approx0.2\,{\rm m}/\,{\rm s}$ is
$u_\ell\,\ell/\nu\approx20,000$, where $\nu\approx10^{-5}\,{\rm m}^2\,{\rm s}^{-1}$
is the viscosity of the gas flow.
Such a large Reynolds number is just within reach on current
supercomputers, but larger domains would remain out of reach for a
long time.
Several earlier works investigated condensational growth of cloud droplets
using Lagrangian tracking in DNS \citep{Paoli09,2015_Sardina,Lozar16,Lanotte09},
but those neglected the collectional growth and only proposed to study the
collectional growth in future work.
An intermediate approach involves the use of Lagrangian
``superparticles'' \citep{Joh12,1997_pruppacher,Shima09,Dullemond_2008},
which represent a ``swarm'' of particles of certain size and number density.
Depending on the values of particle size and number density, there is a
certain probability that an encounter between two superparticles leads to
collectional growth of some of the particles in each swarm (or superparticle).
This superparticle approach has been applied in a recent
LES model to represent the cloud microphysical condensation \citep{Andrejczuk08}
and collection \citep{Andrejczuk10,Riechelmann12,Naumann15} processes.

The purpose of the present paper is to compare the Eulerian approach
involving the Smoluchowski equation with the Lagrangian superparticle
approach with the aim of identifying a promising DNS scheme for tackling
the bottleneck problem of cloud droplets growth.
This has been done in the astrophysical context
\citep{1990_Ohtsuki,Dullemond_2014}, where the principal
problem with the Eulerian 
approach was emphasized in that it requires high mass bin resolution (MBR) 
to avoid artificial speedup of the growth rate.
Here we also compare with the superdroplet approach of \citet{Shima09}.
The original work on this approach was restricted to the case of vanishing
particle inertia, but
this restriction is not a principal limitation of this scheme,
which is in fact well applicable to the case of finite particle inertia.

\section{Lagrangian and Eulerian approaches}

In the following, we refer to the superparticle or superdroplet approaches
as the {\em swarm model}, where each superparticle represents a swarm
of physical particles.
By contrast, the Eulerian approach is also referred to
as the {\em Smoluchowski model}.
Here we compare the two approaches in the meteorological context
of water droplets using, however, simplifying assumptions such as
constant supersaturation and ideal collection efficiency.
In this paper, we generally refer to particles and superparticles,
which are thus used interchangeably with droplets and superdroplets,
respectively.
We begin with a discussion of the gas flows that are being used
in some of the models.

\subsection{Evolution equations for the gas flow in both approaches}

In all the experiments reported below, where a nonvanishing gas flow is
used, we restrict ourselves to two-dimensional (2-D) flows.
However, we also perform several experiments with no gas flow ($\bm{u}=0$).
In those cases the system is spatially uniform and therefore
zero-dimensional (0-D).
For the swarm model, however, each swarm occupies one grid cell,
so it must be treated in at least one dimension (1-D), although
the results of higher-dimensional swarm models will also be discussed,
and they are computational cheaper because they can take advantage of
better parallelization.
By comparison, the Eulerian models are strictly 0-D when there
is no flow.

\subsubsection{Momentum equation of the gas flow}
To obtain $\mbox{\boldmath $u$}$ at each meshpoint, we solve the usual
Navier-Stokes equation
\begin{equation}
{\partial\mbox{\boldmath $u$}\over\partial t}+\mbox{\boldmath $u$}\cdot{\bm{\nabla}}\mbox{\boldmath $u$}={\bm f}
-\rho^{-1}{\bm{\nabla}} p+\bm{F}(\bm{u}), 
\label{turb}
\end{equation}
where ${\bm f}$ is a forcing term,
$p$ is the gas pressure, $\rho$ is the gas density,
which in turn obeys the continuity equation,
\begin{equation}
{\partial\rho\over\partial t}+{\bm{\nabla}}\cdot(\rho\mbox{\boldmath $u$})=0,
\end{equation}
the viscous force $\bm{F}(\bm{u})$ is given by
\begin{equation}
\bm{F}(\bm{u})=\nu(\nabla^2\bm{u}
+{\textstyle{1\over3}}{\bm{\nabla}}{\bm{\nabla}}\cdot\bm{u}
+2\mbox{\boldmath ${\sf S}$}\cdot{\bm{\nabla}}\ln\rho),
\label{Fofu}
\end{equation}
where ${\sc S}_{ij}={\textstyle\frac{1}{2}}(u_{i,j}+u_{j,i})
-{\textstyle{1\over3}}\delta_{ij}{\bm{\nabla}}\cdot\mbox{\boldmath $u$}$ is
the traceless rate-of-strain tensor and commas denote partial differentiation.
We assume that the gas is isothermal and has constant sound speed $c_{\rm s}$
so that the pressure $p=c_{\rm s}^2\rho$ is proportional to the gas density $\rho$.
Note that gravity has been neglected in equation~\eqref{turb}, but this is not a
principal restriction and can be relaxed once suitable non-periodic
boundary conditions are adopted.
For the relatively small domains that can be handled by DNS, gravity will
nevertheless have only minor effects on the fluid flow for atmospheric 
conditions.

\subsubsection{Straining flow}
To obtain a non-vanishing flow, we apply volume forcing via the term ${\bm f}$.
In the case of a time-independent 2-D divergence-free straining flow,
\begin{equation}
\mbox{\boldmath $u$}_{\rm str}=u_0\,(\sin kx \cos kz, 0, -\cos kx \sin kz),
\end{equation}
we take ${\bm f}=\nu k^2\mbox{\boldmath $u$}_{\rm str}$, where $u_0$ determines
the amplitude and $k$ the wavenumber of the flow. 

\subsubsection{Turbulence}
In the case of a turbulent flow, ${\bm f}$ is delta-correlated
in time and consists of random waves in space \citep{2004_Haugen}.
The flow is characterized by a typical forcing wavenumber
$k_{\rm f}$ ($\sqrt{2}k$ for the straining flow or the average wavenumber from a
narrow band of wavevectors) and the root-mean-square (rms) velocity $u_{\rm rms}$.
As a relevant timescale characterizing such a flow, we define
\begin{equation}
	\tau_{\rm cor}=\left(u_{\rm rms}k_{\rm f}\right)^{-1},
\end{equation}
which is an estimate of the correlation time.
This definition is also used for the straining flow, which is a special
case in that it is time-independent and therefore $\tau_{\rm cor}$ would no
longer characterize the correlation time of the flow, but it would
still be proportional to the turnover time.
A simulation without spatial extent can be adopted to investigate
the statistical convergence properties of the Eulerian model regarding
its computational efficiency.

\subsection{Condensational growth}
The growth of the particle radius $r_i$ by condensation is governed 
by \citep{LV11}
\begin{equation}
{{\rm d} r_i\over{\rm d} t}={Gs\over r_i},
\label{cond_eq}
\end{equation}
where $s$ is the supersaturation and $G$ is the condensation
parameter.
Both $s$ and $G$ are in principle dependent on the flow and the
environmental temperature and pressure
\citep[see Chapter~8 of][]{LV11},
but these dependencies are
here neglected, because it would complicate the comparison of
different numerical schemes even further.
Therefore, the condensational growth is driven by constant water
vapor flux without latent heat release in the present study.
We adopt the value $G=5\times10^{-11}\,{\rm m}^2\,{\rm s}^{-1}$
\citep{Lanotte09}.
The assumed constancy of $s$ also implies that the total liquid
water content is not conserved.

\subsection{The swarm model}
\label{SwarmModel}

The swarm model is a Monte Carlo type approach that handles
particle collections in a swarm of particles in a statistical manner
\citep{Dullemond_2008}.
Each swarm $i$ has a particle number density $n_i$,
and occupies a volume $\delta x^D$, which equals the volume of a
fluid grid cell of size $\delta x$ in $D$ dimensions.
All particles in a given swarm have the same mass, radius, and velocity.
Following the description of \citet{Joh12},
the swarm is transported along with its ``shepherd particle'', which is
also referred to as the corresponding superparticle.
The swarm is treated as a
Lagrangian point-particle, where one solves for the particle position
$\mbox{\boldmath $x$}_i$ via
\begin{equation}
\frac{d\mbox{\boldmath $x$}_i}{dt}=\mbox{\boldmath $V$}_i
\label{dxidt}
\end{equation}
and the velocity via
\begin{equation}
\frac{d\mbox{\boldmath $V$}_i}{dt}=\frac{1}{\tau_i}(\mbox{\boldmath $u$}-\mbox{\boldmath $V$}_i)+\bm{g}
\label{dVidt}
\end{equation}
in the usual way.
Here, $\bm g$ is the gravitational acceleration, 
$\tau_i$ is the particle inertial response or stopping time
of a particle in swarm $i$ and is given by
\begin{equation}
\tau_i=\frac{2\rho_{\rm s} r_i^2}{9\rho\nu_i^{\rm eff}},
\label{stopping_time}
\end{equation}
where $r_i$ is the radius of particles in swarm $i$,
$\rho_{\rm s}$ is the particle solid material density, 
$\rho$ is the density of the gas 
and the effective viscosity is given by \cite{Sullivan94}
\begin{equation}
\nu_i^{\rm eff}=\nu\,(1+0.15\,{\rm Re}_i^{0.687}),
\end{equation}
where $\nu$ is the ordinary (microphysical) fluid viscosity, and
${\rm Re}_i=2r_i|\mbox{\boldmath $u$}-\mbox{\boldmath $V$}_i|/\nu$ is the particle Reynolds number,
which provides a correction factor to the particle stopping time.

A given swarm may only interact with every other swarm within
the same grid cell.
The computational cost associated with such
collections scales as $N_{pg}^2$,
where $N_{pg}$ is the number of swarms within a grid cell, but
this is computationally not prohibitive as long
as $N_{pg}$ is not too large.

We now consider two swarms $i$ and $j$ residing within the same grid cell.
Consider first collections of particles within swarm $j$ with a
particle of swarm $i$.
The inverse mean free path of $i$ in $j$ is given by
\begin{equation}
\lambda_{ij}^{-1}=\sigma_{ij} \, n_j \, E_{ij},
\label{eqn:lambda_ij}
\end{equation}
where $\sigma_{ij}$ is the collectional cross section with
\begin{equation}
\sigma_{ij}=\pi(r_{i}+r_{j})^2,
\end{equation}
and $E_{ij}$ is the collision efficiency, but in the following we assume
$E_{ij}=1$ in all cases.\footnote{
In \cite{Shima09}, the mean free path 
is defined by invoking the swarm with the larger
number density of
physical particles; see section~\ref{CoagII} for details.}
The particle number density in swarm $j$ is $n_j$
and $r_{i}$ and $r_{j}$ represent the radii of the particles in the
two swarms.
From this, one can find the typical rate of collections between a 
particle of swarm $i$ and particles of swarm $j$ as
\begin{equation}
\tau_{ij}^{-1}=\lambda_{ij}^{-1}\left|\mbox{\boldmath $V$}_{i}-\mbox{\boldmath $V$}_{j}\right|
=\sigma_{ij}n_j\left|\mbox{\boldmath $V$}_{i}-\mbox{\boldmath $V$}_{j}\right| E_{ij},
\label{tauij1}
\end{equation}
where $\mbox{\boldmath $V$}_{i}$ and $\mbox{\boldmath $V$}_{j}$ are the velocities of swarms $i$ and $j$.
The probability of a collection between the swarm $i$ and
any of the particles of swarm $j$ within the current time step
$\Delta t$ is then given by
\begin{equation}
\label{prob}
p_{ij}=\tau_{ij}^{-1}\Delta t.
\end{equation}
This effectively puts a restriction on the time step,
since the probability cannot be larger than unity.
For each swarm pair in a grid cell, one now picks a random number,
$\eta_{ij}$, and compares it with $p_{ij}$.
A collection event occurs in the case when $\eta_{ij} < p_{ij}$.

\subsubsection{Collection scheme I}
\label{CoagI}

For the swarm model, two different collection schemes have been proposed
in the astrophysical and meteorological contexts.
We begin discussing the former (scheme~I), which is similar to that
described by \citet{Joh12} in that it maintains a constant mass of the
individual swarms.
In the context of mathematical probability, this approach is also
known as mass flow algorithm \citep{EW01,PWK11}.
Scheme~II is discussed in section~\ref{CoagII}.

If $\eta_{ij}<p_{ij}$, one assumes that \emph{all} the particles in swarm $i$
have collided with a particle in swarm $j$. In this collection scheme, 
all swarms are treated individually. This means that even though the 
particles in swarm $i$ have
collided with the particles in swarm $j$, swarm $j$ is kept unchanged at this
stage. Instead, swarm $j$ is treated individually at a different stage.
Hence, all collections are asymmetric, i.e., $p_{ij}\ne p_{ji}$.
The new mass of the particles in swarm $i$ now becomes
\begin{equation}
\tilde{m}_i=m_{i}+m_j,
\end{equation}
where $m_{i}$ is the mass before the collection and the tilde represents
the new value after collection.
In order to ensure mass conservation, the {\em total} mass of swarm $i$
is kept unchanged, i.e.,
\begin{equation}
\tilde{n}_{i}\tilde{m}_i=n_{i}m_{i},
\end{equation}
which implies that the new particle number density, $\tilde{n}_i$,
is given by $\tilde{n}_{i}=n_{i}m_{i}/\tilde{m}_i$; see equation~(17) of
\cite{PWK11} for the corresponding treatment in the mass flow algorithm.
By invoking momentum conservation,
\begin{equation}
\tilde{\mbox{\boldmath $V$}}_{i}\tilde{m}_{i}=\mbox{\boldmath $V$}_{i}m_{i}+\mbox{\boldmath $V$}_{j}m_j,
\end{equation}
the new velocity of any particle in swarm $i$ is given by
$\tilde{\mbox{\boldmath $V$}}_{i}=(\mbox{\boldmath $V$}_{i}m_{i}+\mbox{\boldmath $V$}_{j}m_j)/\tilde{m}_{i}$.

\subsubsection{Collection scheme II}
\label{CoagII}

In the meteorological context, the following collection scheme has been
proposed \citep{Shima09}.
Assume two swarms $i$ and $j$, and consider (without loss of generality)
the case $n_j>n_i$.
The collection probability
of particles in swarm $i$ with swarm $j$ is, again, given by equation~\eqref{prob}.
If the two swarms are found to collide, the new masses of the particles in
the two swarms are given by
\begin{eqnarray}
\tilde{m}_i&=&m_{i}+m_j, \nonumber \\
\tilde{m}_j&=&m_j,
\end{eqnarray}
but now their new particle number densities are
\begin{eqnarray}
\tilde{n}_i&=&n_i, \nonumber \\
\tilde{n}_j&=&n_j-n_i.
\end{eqnarray}
In other words, the number of particles in the smaller swarm
remains unchanged (and their masses are increased), while that in
the larger one is reduced by the amount of particles that
have collided with all the particles of the smaller swarm
(and their masses remain unchanged).
This implies that in equation~\eqref{eqn:lambda_ij},
the mean free path is defined with respect to the swarm with the larger
number density of physical particles, as explained in \cite{Shima09}.
Finally, the new momenta of the particles in the two swarms are given by
\begin{eqnarray}
\tilde{\mbox{\boldmath $V$}}_{i}\tilde{m}_{i}&=&\mbox{\boldmath $V$}_{i}m_{i}+\mbox{\boldmath $V$}_{j}m_j, \nonumber \\
\tilde{\mbox{\boldmath $V$}}_{j}\tilde{m}_{j}&=&\mbox{\boldmath $V$}_{j}m_j.
\end{eqnarray}
In contrast to scheme~I, these collections are symmetric, i.e.\
$p_{ij} = p_{ji}$.
Consequently, both swarms are changed during a collection.
However, the asymmetric collection property of scheme~I of \citep{Joh12}
may not have been previously recognized, nor has its accuracy been
compared with other models, which we will further discuss below.

\subsubsection{Initial particle distribution}

We recall that particles within a swarm may interact with particles of
another swarm only if both swarms occupy the same grid cell.
The effective volume of each swarm is therefore equal to $\delta x^D$,
where $D$ is the spatial dimension introduced in section~\ref{SwarmModel}.
The total number of particles in our computational domain is therefore
$\delta x^D$ times the sum of $n_i$ over all $N_p$ swarms.
This must also be equal to $n L^D $, where $n$ is the total number density
represented by the simulation and $L$ is the size of the computational domain.
Thus, we have
\begin{equation}
n L^D=\delta x^D \sum_{i=1}^{N_p}n_i.
\end{equation}
Initially ($t=0$), the particle number densities of
all swarms are the same and since $(L/\delta x)^D=N_{\rm grid}$ is the
total number of grid points, we have $n N_{\rm grid}=n_i N_p$.
Thus, the initial number density of particles within one swarm must be
\begin{equation}
n_i=n N_{\rm grid}/N_p\quad(\mbox{at $t=0$}).
\end{equation}
In the following, we choose the initial particle size distribution of total
physical particles in the domain to be log-normal, i.e.,
\begin{equation}
\label{init_dist}
f(r_i,0)=\left(n_0\left/(\sqrt{2\pi}\sigma_{\rm p}r)\right.\right)\exp\left\{
-[\ln (r_i/r_{\rm ini})]^2/2\sigma_{\rm p}^2\right\},
\end{equation}
where $r_{\rm ini}$ and $\sigma_{\rm p}$ are the center and width of
the size distribution, respectively; $n_0=n(t=0)$ is the initial total
number density of physical particles.
These particles are distributed uniformly over 
all swarms within the computational domain.
This means that particles in each swarm are of the same size,
but different from swarm to swarm.

\subsection{Eulerian approach}
\label{eul_app}

To model the combined growth of particles through condensation
and collection in a multi-dimensional flow in the Eulerian description,
we describe the evolution of particles of different radii $r$
(or, equivalently, of different logarithmic particle mass $\ln m$)
at different positions $\mbox{\boldmath $x$}$ and time $t$.
We employ the particle
distribution function $f(\mbox{\boldmath $x$},r,t)$, or, alternatively in terms of
logarithmic particle mass $\ln m$, $\tilde{f}(\mbox{\boldmath $x$},\ln m,t)$,
such that the total number density of particles is given by
\begin{equation}
n(\mbox{\boldmath $x$},t)=\int_0^\infty f(\mbox{\boldmath $x$},r,t)\, {\rm d} r,
\end{equation}
or, correspondingly for $\tilde{f}$, we have
$n(\mbox{\boldmath $x$},t)=\int_{-\infty}^\infty \tilde{f}(\mbox{\boldmath $x$},\ln m,t)\, {\rm d}\ln m$.
Since $m=4\pi r^3\rho_{\rm s}/3$, we have
$\tilde{f}=f\,{\rm d} r/{\rm d}\ln m=fr/3$.
Note that $n(\mbox{\boldmath $x$},t)$ obeys the usual continuity equation,
\begin{equation}
{\partial n\over\partial t}+{\bm{\nabla}}\cdot(n\overline{\bm{v}})=D_{\rm p}\nabla^2n,
\label{eqn:dndt}
\end{equation}
where $\overline{\bm{v}}$ is the mean particle velocity
(i.e., an average over all particle sizes) and
$D_{\rm p}$ is a Brownian diffusion term, which is enhanced for
numerical stability and will be chosen depending on the mesh resolution.
The evolution of the particle distribution function is governed by a
similar equation, but with additional coupling terms due to condensation
and collection, i.e.
\begin{equation}
{\partial f\over\partial t}+{\bm{\nabla}}\cdot(f\mbox{\boldmath $v$})
+\nabla_r(f C)={\cal T}_{\rm coll}+D_{\rm p}\nabla^2f,
\label{eqn:dfdt}
\end{equation}
where $\nabla_r=\partial/\partial r$ is the derivative with respect to $r$,
$C\equiv{\rm d} r/{\rm d} t=Gs/r$, as given in equation~\eqref{cond_eq},
and ${\cal T}_{\rm coll}$ describes the change of the number density
of particles for smaller and larger radii, as will be defined below.
Furthermore, $\mbox{\boldmath $v$}(\mbox{\boldmath $x$},r,t)$ is the particle velocity within the
resolved grid cell, which is discussed below.
It also determines the mean flow $\overline{\bm{v}}=\int f\bm{v}\,{\rm d}r/n$
in equation~\eqref{eqn:dndt}.

The modeling of condensation and collection implies coupling of
the evolution equations of $f(\mbox{\boldmath $x$},r,t)$ for different values of $r$.
The advantage of using $\tilde{f}(\mbox{\boldmath $x$},\ln m,t)$ is that it allows us to
cover a large range in $m$, because we will use then an exponentially
stretched grid in $m$ such that $\ln m$ is uniformly spaced \citep{1997_pruppacher,SY01,Joh04}.
The total number density within a finite mass interval $\delta\ln m$ is
then given by $\tilde{f}(\mbox{\boldmath $x$},\ln m,t)\,\delta\ln m$.
Thus, the total number density of particles of all sizes at position $\mbox{\boldmath $x$}$
and time $t$ is given by
\begin{equation}
\label{n_ftilde}
n(\mbox{\boldmath $x$},t)=\sum_{k=1}^{k_{\max}} \tilde{f}_k \, \delta\ln m
=\sum_{k=1}^{k_{\max}} \hat{f}_k,
\end{equation}
where $\hat{f}_k=\tilde{f}(\ln m_k) \, \delta\ln m$
is the variable used in the simulations
and $k_{\max}$ is the number of logarithmic mass bins.

Let us first consider the process of condensation, which is described
in equation~\eqref{eqn:dfdt} by the term $\nabla_r(f C)$, where $f C$
is the flux of particle from one size bin to the next.
Evidently, the total number density is only conserved if the particle
flux $fC$ vanishes for $r=r_{\min}$ and $r=r_{\max}$, which is the case
if the range of $r$ is sufficiently large.
In particular, $(fC)_{\min}\to0$, because $n\to0$ for $m\to0$.
In practice, however, we consider finite lower cutoff values of $m$
and therefore expect some degree of mass loss at the smallest
mass bins.
The same is also true for the largest mass bin once the size distribution
has grown to sufficiently large values.
In all cases with pure condensation, it is convenient to display
solutions in non-dimensional form by measuring time in units of
\begin{equation}
\tau_{\rm cond}=r_{\rm ini}^2/2Gs
\label{eqn:tau}
\end{equation}
and $r$ in units of $r_{\rm ini}$. 
We refer to Appendix~\ref{AppA} for more details on the
condensation equation for the Eulerian approach.

Next, we consider collection, which leads to a decrease of $n$, but
does not change the mean mass density of liquid water.
The evolution of $\tilde{f}(\mbox{\boldmath $x$},\ln m,t)$ due to collection
is governed by the Smoluchowski equation
\begin{eqnarray}
{\cal T}_{\rm coll}={\textstyle\frac{1}{2}}\!\!\int_0^m \!\!\! K(m-m',m')\,f(m-m')\,f(m')\,{\rm d} m'\nonumber\\
-\int_0^\infty K(m,m')\,f(m)\,f(m')\,{\rm d} m'.
\label{Tcoag}
\end{eqnarray}
Here, $K$ is a kernel, which is proportional to the collision efficiency
$E(m,m')$ and a geometric contribution.
As mentioned above, we assume $E=1$ and so $K$ is given by
\begin{equation}
K(m,m')=\pi(r+r')^2|\mbox{\boldmath $v$}-\mbox{\boldmath $v$}'|,
\end{equation}
where $r$ and $r'$ are the radii of the corresponding mass variables,
$m$ and $m'$, while $\mbox{\boldmath $v$}$ and
$\mbox{\boldmath $v$}'$ are their respective velocities,
whose governing equation is given below.

In the following, we define the mass and radius bins such that
\begin{equation}
m_k=m_1\delta^{k-1},\quad
r_k=r_1\delta^{(k-1)/3}.
\end{equation}
Unfortunately, $\delta=2$ is in many cases far too coarse, so we take
\begin{equation}
\delta=2^{1/\beta},
\label{delta_expr}
\end{equation}
where $\beta$ is a parameter that we chose to be a power of two.
For a fixed mass bin range, the number of mass bins $k_{\rm max}$
increases with increasing $\beta$.
In terms of $\hat{f}_k$, equation~\eqref{Tcoag} reads
\begin{equation}
{\cal T}^{\rm coll}_k={\textstyle\frac{1}{2}}\!\!\sum_{i\,+j\,\in\, k}
K_{ij} \, {m_i+m_j\over m_k} \, \hat{f}_i\hat{f}_j
-\hat{f}_k\sum_{i=1}^{k_{\max}} K_{ik}\hat{f}_i,
\label{Tcoag2}
\end{equation}
where we have adopted the nomenclature of \citet{Joh04}, where
$i\,+j\,\in\, k$ denotes all values of $i$ and $j$ for which
\begin{equation}
m_{k-1/2}\leq m_i+m_j<m_{k+1/2}
\label{massIntervl}
\end{equation}
is fulfilled.
The term $(m_i+m_j)/m_k$ in equation~\eqref{Tcoag2} comes
from the fact that collections between cloud droplets from two mass bins
may not necessarily result in a cloud droplet mass being exactly in
the middle of the nearest mass bin.
\cite{Joh04} therefore included this factor so that mass is strictly conserved.
The discrete kernel is then $K_{ij}=\pi(r_i+r_j)^2|\mbox{\boldmath $v$}_i-\mbox{\boldmath $v$}_j|$.

The corresponding momentum equations for the velocities
$\mbox{\boldmath $v$}_k(\mbox{\boldmath $x$},t)=\mbox{\boldmath $v$}(\mbox{\boldmath $x$},\ln m_k,t)$ for each logarithmic
mass value $\ln m_k$ is
\begin{equation}
{\partial\mbox{\boldmath $v$}_k\over\partial t}+\mbox{\boldmath $v$}_k\cdot{\bm{\nabla}}\mbox{\boldmath $v$}_k
=\bm{g}-{1\over\tau_k}(\bm{v}_k-\bm{u})
+\bm{F}_k(\bm{v}_k)+{\cal M}_k
,\quad 
1\leq k\leq k_{\max}.
\label{dvdt}
\end{equation}
Here, $\mbox{\boldmath $u$}$ is the gas velocity,
$\tau_k$ (for $k=i$) is defined by equation~\eqref{stopping_time},
and 
\begin{equation}
\bm{F}_k(\bm{v}_k)=\nu_p\nabla^2\bm{v}_k
\label{particle_viscous}
\end{equation}
is a viscous force among particles, which
should be very small for dilute particle suspensions,
but is nevertheless retained in equation~\eqref{dvdt} for the sake of 
numerical stability of the code.
It is not to be confused with the drag force,
$-\tau_k^{-1}(\bm{v}_k-\bm{u})$ between particles and gas.
In principle, the expression for $\bm{F}_k(\bm{v}_k)$ should be based on
the divergence of the traceless rate-of-strain tensor of $\mbox{\boldmath $v$}_k$,
similarly to the corresponding expression for the viscous force of the
gas discussed in equation~\eqref{Fofu}.
However, since the term $\bm{F}_k(\bm{v}_k)$ is unphysical anyway, we just
use the simpler expression proportional to $\nabla^2\bm{v}_k$ instead.

The linear momentum of all particles is given by
$\sum\langle\hat{f}_k m_k\bm{v}_k\rangle$,
where angle brackets denote volume averages.
In order that this quantity is conserved by each collection,
the target has to receive a corresponding kick, which leads
to the last term in equation~\eqref{dvdt}, but it leaves
the velocities of the collection partners unchanged.
It is therefore only related to the first term on the
right-hand side of equation~\eqref{Tcoag2} and not
the second, so it is given by (see Appendix~\ref{momentumC})
\begin{equation}
{\cal M}_k=\frac{1}{2\hat{f}_km_k}
\!\!\,\,\,\,\sum_{i\,+j\,\in\, k} K_{ij}\hat{f}_i\hat{f}_j\,\left[
m_i\mbox{\boldmath $v$}_i+m_j\mbox{\boldmath $v$}_j-(m_i+m_j)
\mbox{\boldmath $v$}_k\right].
\label{Mkterm}
\end{equation}
To our knowledge, this momentum-conserving term has not been included
in any of the very few earlier works that include a momentum equation
for each particle species \citep[cf.][]{SY01,Elperin_etal15}.
The reason why this has apparently not previously been discussed in 
the literature is that in meteorological applications one usually works
with the averaged kernel and neglects the evolution of the velocities
for the different mass bins \citep{Grabowski_2013}.
This correction term is evidently zero when the momentum
of the two collection constituents ($=m_i\bm{v}_i+m_j\bm{v}_j$)
is equal to that of the resulting constituent [$=(m_i+m_j)\bm{v}_k$].
Nevertheless, as is shown in Appendix~\ref{momentumC}, the momentum
conserving correction changes the time evolution of the droplet spectrum
in an unexpected way when the MBR is high, but the results are similar
for $\beta=2$.
Furthermore, for turbulent flows, as is discussed below, these correction
terms become insignificant.

As mentioned above, a shortcoming of the Eulerian approach is that
no collection is possible from equally sized particles.
To assess the consequences of this unphysical limitation,
we study the sensitivity of the results to replacing
$K_{i\,i}$ either (i) by $(K_{i+1\;i}+K_{i\;i+1})/2$ or (ii) by
$\epsilon_{\rm self}\pi(2r_i)^2|\mbox{\boldmath $v$}_i+\mbox{\boldmath $v$}_j|/2$, where
$\epsilon_{\rm self}$ is an empirical parameter.

\subsection{Boundary conditions and diagnostics}

In the present work, we use periodic boundary conditions for all variables
in all directions.
Therefore, no particles and no gas are lost through the boundaries
of the domain.
This approximation is reasonable as long as we are
interested in modeling a small domain well within a cloud where also
heavier particles can be assumed to enter from above.
The use of periodic boundary conditions requires us to neglect
gravity in equation~\eqref{turb}, which could be relaxed if non-periodic
boundary conditions were adopted.

To characterize the size distribution, especially for the larger particles,
we consider the evolution of different normalized moments of the size spectra,
\begin{equation}
a_\zeta=\left(\sum_{k=1}^{k_{\max}} \left\langle\hat{f}_k\,r_k^\zeta\right\rangle\left/\,
\sum_{k=1}^{k_{\max}} \left\langle\hat{f}_k\right\rangle\right. \right)^{1/\zeta},
\label{azeta}
\end{equation}
where $\zeta$ is a positive integer.
The mean radius $\overline{r}$ is given by $a_1$.
Higher moments represent the tail of the distribution at large radii.
In view of raindrop formation, we will be particularly
interested in the largest droplets in the distribution.
However, very large moments become numerically difficult to compute
accurately, but $a_{12}$, for example, was still
not sufficiently representative of the largest droplets.
Therefore we arrived at $a_{24}$ as a reasonable compromise to characterize
the largest droplets in the distribution.
Alternatively, the size distribution can be characterized by a gamma
distribution, which requires the determination of only three moments
in an approach known as the three-moment bulk scheme \citep{Seifert2001}.
This will be discussed in more detail in section~\ref{Characterizing}.

In the case of collection, the condensation timescale $\tau_{\rm cond}$,
defined in equation~\eqref{eqn:tau}, is no longer relevant, but it is instead a
collection timescale that can be defined in the Eulerian model as
\begin{equation}
\tau_{\rm coll}^{-1}=\left.\sum_{k=1}^{k_{\max}}{\left\langle {{\cal T}_k^{\rm coll}}\right\rangle}
\right/ \sum_{k=1}^{k_{\max}}{\left\langle {\hat{f}_k}\right\rangle},
\end{equation}
which is, in this definition, a time-dependent quantity.
In the Lagrangian model, this quantity can be defined by 
the collection frequency.
Unlike the case of pure condensation, where $\tau_{\rm cond}$
is the appropriate time unit, $\tau_{\rm coll}$ can
only be used a posteriori as a diagnostic quantity.
However, given that the speed of pure collection is proportional
to the mean particle density $n$, it is often convenient to perform
simulations at increased values of $n$ and then rescale time to
a fixed reference density $n_{\rm ref}$ and use
\begin{equation}
\tilde{t}=t\,n_0/n_{\rm ref}.
\label{ttilde}
\end{equation}
In the following we use $n_{\rm ref}=10^8\,{\rm m}^{-3}$, which is the typical
value of $n$ in atmospheric clouds.
Analogously we also define $\tilde{\tau}_{\rm coll}=\tau_{\rm coll}\,n_0/n_{\rm ref}.$
Finally, the number of particles in the total simulation
domain is $N(t)=\int n(\bm{x},t)\,{\rm d}^Dx$.

\subsection{Computational implementation}
\label{imple}

We use the {\sc Pencil Code}\footnote{https://github.com/pencil-code/},
which is a public domain code where
the relevant equations have been implemented \citep{Joh04,JAB04,Bab15}.
We refer to Appendix~\ref{AppA} for a description of an important modification
applied to the implementation of equation~\eqref{cond_eq}.
The implementation of equation~\eqref{Tcoag2} has been discussed in detail by
\cite{Joh04}, and follows an approach described earlier by \citet{SY01}.
However, momentum conservation during collections
was previously ignored in the Eulerian model.
The current revision number is 73563 when checking out the code via the
svn bridge on the public github repository.

When traditional point particle Lagrangian particle tracking is employed,
it is usually beneficial to employ higher order interpolation between the
neighboring grid cells to find the value of a given fluid variable at
the exact position of the particle.
By default, the cloud-in-cell (CIC) algorithm is used, which involves
first order interpolation for the particle properties on the mesh.
In the swarm approach, however, the particles in each swarm fills
the volume of a grid cell in which the shepherd particle is.
The distribution of the swarm throughout the grid cell is homogeneous and
isotropic, and as such the swarm has no particular position within the grid cell.
It is true that there is a particular position associated with the
swarm, namely the position of the shepherd particle, but this position 
has no purpose other than to determine in which grid cell the swarm resides.
Below we shall show that it is {\emph not} better to use any kind
of interpolation in determining the value of the fluid variables at the
position of the swarm, but rather to use the values of the grid cell in
which the swarm resides.
This method is technically referred to as nearest grid point mapping (NGP).
Details concerning each experiment are summarized in Table~\ref{runs}.

\section{Results}

\subsection{Condensation experiments}
\label{CondensationExperiments}

\begin{table}[t!]
\caption{Summary of the simulations.}
\centering
\setlength{\tabcolsep}{3pt}
\begin{tabular}{l c c c c c c c c c c c p{1.5cm}}
\hline
Run & Scheme & Dim & $L$ (m) & $N_p$ & $N_{\rm grid}$ & IM & Processes &
$\beta$ & $n_0$ (${\rm m^{-3}}$) & Flow & $D_p$ (${\rm m^2}/{\rm s}$) & $\nu_p$ (${\rm m^2}/{\rm s}$) \\
\hline
1A &SwI & 3-D & $0.5$ & $10^4$ &  $16^3$ &CIC& Con & -- & $10^{10}$ & -- \\ 
2B &Eu & 0-D & $0.5$ & -- & -- & -- & Con & $128$ & $10^{11}$ & -- \\
3C &Eu & 0-D & $0.5$ & -- & -- & -- & Col & $128$ & $10^{11}$ & grav \\
4B &SwI & 3-D & $0.5$ & $32N_{\rm grid}$ &  $32^3$ &CIC& Col & -- & $10^{10}$ & grav \\ 
6A &SwII & 3-D & $0.5$ & $32N_{\rm grid}$ &  $32^3$ &CIC& Col & -- & $10^{10}$ & grav \\ 
7A &SwII & 2-D & $2\pi$ & $3\times10^5$ &  $64^2$ &CIC& Col & -- & $10^{10}$ & strain \\ 
7B &SwII & 2-D & $2\pi$ & $3\times10^5$ &  $128^2$ &CIC& Col & -- & $10^{10}$ & strain \\ 
7C &SwII & 2-D & $2\pi$ & $3\times10^5$ &  $256^2$ &CIC& Col  & -- & $10^{10}$ & strain \\ 
7D &Eu & 2-D & $2\pi$ & -- & $128^2$ &--& Col  & $2$ & $10^{10}$ & strain & $0.05$ & $0.01$ \\ 
7E &SwII & 2-D & $2\pi$ & $3\times10^5$ &  $80^2$ &NGP& Col & -- & $10^{10}$ & strain \\ 
7F &SwII & 2-D & $2\pi$ & $3\times10^5$ &  $160^2$ &NGP& Col & -- & $10^{10}$ & strain \\ 
9A &SwII & 2-D & $2\pi$ & $5\times10^4$ &  $128^2$ &CIC& Both & -- & $10^8$ & strain \\ 
9B &SwII & 2-D & $2\pi$ & $5\times10^4$ &  $128^2$ &NGP& Both & -- & $10^8$ & strain \\ 
9C &Eu & 2-D & $2\pi$ & -- & $128^2$ &--& Both & $2$ & $10^8$ & strain & $0.02$ & $0.10$ \\ 
9D &Eu & 2-D & $2\pi$ & -- & $128^2$ &--& Both & $2$ & $10^8$ & strain & $0.01$ & $0.05$ \\ 
9E &Eu & 2-D & $2\pi$ & -- & $256^2$ &--& Both & $2$ & $10^8$ & strain & $0.005$ & $0.05$ \\ 
10A &Eu & 2-D & $0.5$ & -- & $512^2$ &--& Col  & $2$ & $10^{10}$ & turb & $0.001$ & $0.001$ \\ 
10B &SwII & 2-D & $0.5$ & $1.2\times10^6$ &  $512^2$ &NGP& Col & -- & $10^{10}$ & turb \\ 
\hline
\multicolumn{13}{p{\textwidth}}{``IM'' denotes the interpolation method, 
``Col'' refers to collection, 
``Con'' refers to condensation, ``Eu'' refers to Eulerian model, 
``SwI'' refers to collection scheme I of swarm model, 
``SwII'' refers to collection scheme II of swarm model,
``Both'' refers to condensation
and collection, ``grav'' refers to gravity ($\bm{u}$=0), ``strain'' refers to
straining flow, ``turb'' refers to turbulence, and ``Dim'' refers to the dimension.
}\end{tabular}
\label{runs}
\end{table}

\begin{figure}[t!]\begin{center}
\includegraphics[width=\textwidth]{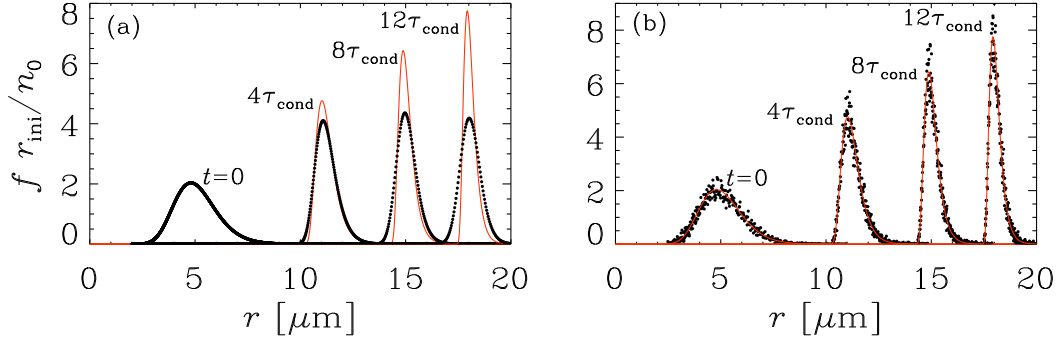}
\end{center}\caption[]{
Comparison of the numerically obtained size spectra with the analytic solution
for condensation with a lognormal initial condition given by
$r_{\rm ini}=5\,\mu{\rm m}$, and $\sigma_{\rm p}=0.2$.
Simulations of pure condensation (no turbulence nor gravity) with the
Eulerian model (a) using $\beta$=128 and $k_{\rm max}$=1281 mass bins in the range
$2$--$20\,\mu{\rm m}$ and the Lagrangian swarm model (b) with
$N_p=10000$ and $N_{\rm grid}=16^3$.
The solid lines correspond to the analytic solution given by
equation~\eqref{cond_seinfeld} while the black dots represent the numerical results.
See run 1A and 2B of Table~\ref{runs} for simulation details.}
\label{psol_comp_5em6}
\end{figure}

We compare the Eulerian and Lagrangian models for
the pure condensation process without motion, i.e., zero gas velocity.
In the case of homogeneous condensation, we can compare the numerical
solution with the analytic solution of \citet{Sei06}; see their Fig.~13.25.
To this end, we make use of the fact that solutions of the condensation
equation~\eqref{cond_eq} obey
\begin{equation}
f(r,t)=(r/\tilde{r})\,f(\tilde{r},0),
\end{equation}
where $\tilde{r}$ is a shifted coordinate with $\tilde{r}^2=r^2-2Gst$.
With the log-normal initial distribution given by equation~\eqref{init_dist}, this
yields
\begin{equation}
\label{cond_seinfeld}
f(r,t)= \frac{n_0}{\sqrt{2\pi}\,\sigma_{\rm p}}{r\over{\tilde{r}^2}}
\exp\left[-{(\ln\tilde{r}-\ln r_{\rm ini})^2
\over 2\sigma_{\rm p}^2} \right],
\end{equation}
where $r_{\rm ini}$ denotes the position of the peak of the distribution and
$\sigma_{\rm p}=\ln\sigma_{\rm SP}$ denotes its width, where $\sigma_{\rm SP}$
is the symbol introduced by \citet{Sei06}.
What is remarkable here is the fact that $f(r,t)$ vanishes for
$r<r_\ast\equiv\sqrt{2Gst}$.
This is because in this model, no new particles
are created and even particles of zero initial radius will have
grown to a radius $r_\ast$ after time $t$.
Furthermore, the small particles with $r=r_\ast$ grow faster
than any of the larger ones, which leads to a sharp rise
in the distribution function at $r=r_\ast$.
Thus, $\partial f/\partial r$ has a discontinuity at $r=r_\ast$.
This poses a challenge for the Eulerian scheme in which the derivative
$\partial/\partial r$ is discretized; see equation~\eqref{eqn:dfdt}.
In Figure~\ref{psol_comp_5em6}, we compare solutions obtained using both Eulerian
and Lagrangian approaches.
It is evident that the $r$-dependence obtained from the
Eulerian solution is too smooth compared with the
analytic one, even though we have used 1281 mass bins
with $\beta$=128 to represent $r$
on our logarithmically spaced mesh over the range
$2\,\mu{\rm m}\leq r\leq20\,\mu{\rm m}$, which corresponds
to $\delta\approx1.0054$; see equation~\eqref{delta_expr}.
Better accuracy could be obtained by using a uniformly spaced grid in $r$,
but this would not be useful later when the purpose is to consider
collection spanning a range of several orders of magnitude in radius.
By comparison, the Lagrangian solution shown in the right-hand panel of
Figure~\ref{psol_comp_5em6} (here with $n_0=10^{10}\,{\rm m}^{-3}$) has no difficulty in
reproducing the discontinuity in $\partial f/\partial r$ at $r=r_\ast$.
Moreover, the Lagrangian solution agrees perfectly with the analytical solution.

In practice, for the Eulerian approach we would use logarithmic
spacing on a mesh with $\delta=2$
or $2^{1/2}\approx1.414$.
However, in such cases, the distribution develops a broad tail.
This is demonstrated in detail in Appendix~\ref{MBR_dependency}.
On the other hand, as we show further below, for turbulent and other
velocity fields, the results depend much less on MBR so that computations
with $\beta=2$ can be sufficiently accurate.

\subsection{Purely gravitational collection experiments}

We now consider uniform collection with no spatial
variation of the velocity and density fields for both the gas and the particles.
For the purely geometrical kernel, no analytic solution exists.
However, we can compare the convergence properties of our two quite
different numerical approaches and thereby get some sense of their
validity in cases when the two agree.
We consider pure collection experiments, starting again with a
log-normal distribution.
The results are presented in terms of normalized time;
see equation~\eqref{ttilde}.

\begin{figure}[t!]\begin{center}
\includegraphics[width=\textwidth]{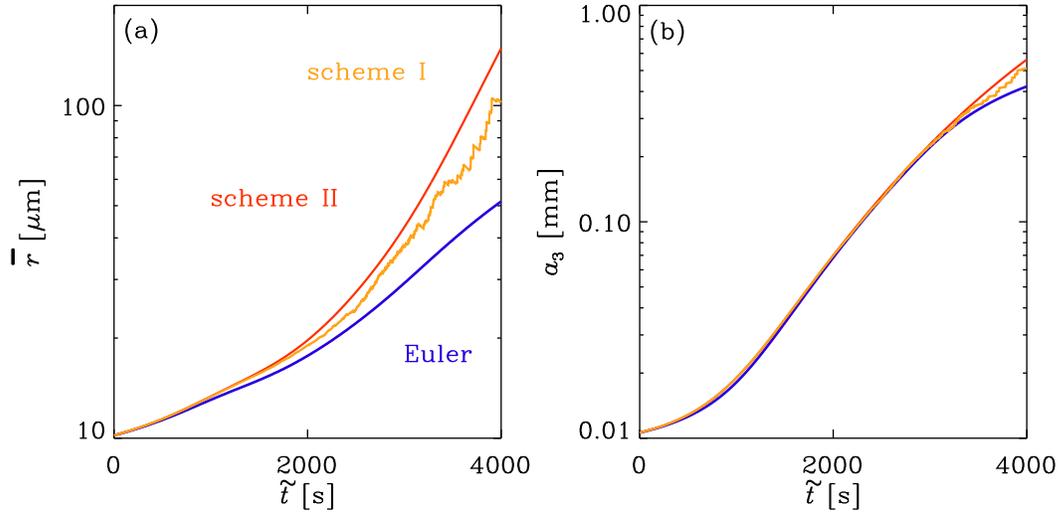}
\end{center}\caption[]{3-D simulations with the swarm model
and $32^3$ grid points using schemes~II (red) and I (orange),
compared with the Eulerian model with $\beta=128$ (solid blue line) for
(a) $\bar{r}$ and (b) $a_3$ . The
collection is driven by gravity.
See Runs~3C, 4B, and 6A of Table~\ref{runs} for simulation
details.}
\label{moments_swarm_Euler}
\end{figure}

\subsubsection{Comparison between swarm collection schemes I and II}

In Figure~\ref{moments_swarm_Euler}, we compare
schemes~I and II of the swarm model together with the Eulerian model.
The simulations have been performed with $N_{\rm grid}=32^3$ grid points
and $N_p=32N_{\rm grid}$ swarms 
(the statistics is converged for $N_p/N_{\rm grid} \ge 4$, 
as discussed in Appendix~\ref{swarm_statistics}).
The left-hand panel of Figure~\ref{moments_swarm_Euler} shows
that for $\bar{r}$ the results of the swarm simulations 
with scheme~I agree with those of scheme~II at
early times, but depart at late times.
However, for $a_3$, the agreement is excellent,
as shown in the right-hand panel of Figure~\ref{moments_swarm_Euler}.
The evolution of $\bar{r}$ with scheme~I shows considerable scatter at late times.
We recall that the main difference between schemes~I and
II is the geometry of collections.
The collections simulated with scheme~I are asymmetric,
while those with scheme~II are symmetric.
Thus, in scheme~II both swarms change either their total mass or their
total particle number, while in scheme~I the total mass of a swarm
is kept constant by adjusting the particle number correspondingly.
This property of scheme~I may be responsible for creating stronger
fluctuations in the mean radius.
Therefore, to keep the amount of scatter comparable, scheme~II is
effectively less demanding.
In the following, we will mainly 
adopt scheme~II to save computational time.

\subsubsection{Comparison between collection scheme II and the Eulerian model}

As we have seen above, the
swarm simulations follow the 
Eulerian results rather well for $a_3$ (see
the right-hand panel of Figure~\ref{moments_swarm_Euler}), but are somewhat
different for $\bar{r}$.
At early times, on the other hand, the evolution of $\overline{r}$ obtained
with the swarm model with collection scheme~I follows more closely that
of the Eulerian model.
However, at later times, the evolution of $\overline{r}$ obtained
with the swarm model departs from the one simulated with the Eulerian model.
This is surprising and might hint at a false convergence behavior,
especially of the swarm model which has very few particles at small radii.
This interpretation is supported by the fact that at larger radii the
agreement is better, which also is the physically more relevant case.

We show in Appendix~\ref{MBR_dependency_coag} that, in the case of purely
gravity-driven collections, $\bar{r}$ converges only for very large MBR.
Thus, the MBR dependency of the numerical solution using the Smoluchowski
scheme appears to be a serious obstacle in studying particle growth not
only by condensation, but also by collection.
This is a strong argument in favor of the Lagrangian scheme.
The evolution of $a_3$, on the other hand, agrees rather well between
the swarm and Eulerian models.

\begin{figure}[t!]\begin{center}
\includegraphics[width=.8\textwidth]{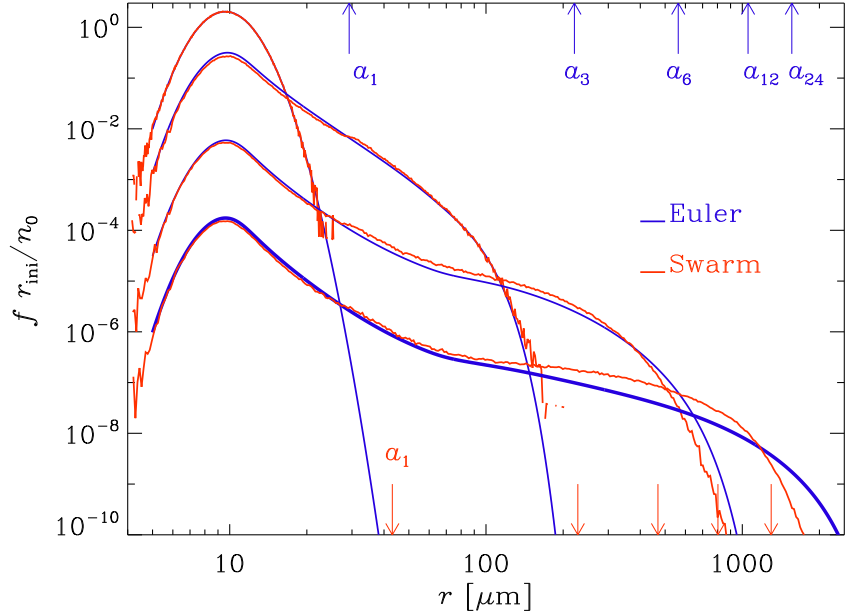}
\end{center}\caption[]{Same simulations as in 
Figure~\ref{moments_swarm_Euler}, but here we
only compare scheme~II with the Eulerian model.
Size spectra are given for $\tilde{t}$=$0\,{\rm s}$, $1000\,{\rm s}$,
$2000\,{\rm s}$ and $3000\,{\rm s}$. 
The arrows show the values of $a_1$, $a_3$, $a_6$, $a_{12}$,
and $a_{24}$ for $\tilde{t}$=$3000\,{\rm s}$.
}\label{moments_shima_Nx32}
\end{figure}

We emphasize that $\overline{r}$ is sensitive to subtle changes in the
size distribution, but it is at the same time not really relevant to
characterizing the collectional growth toward large particles.
As is shown in the following sections, the mean particle radius
often increases by not much more than a factor of three (see also the
left-hand panel of Figure~\ref{moments_swarm_Euler}), while the size
distribution can become rather broad and its tail can reach the size
of raindrops within a relatively short time.
In addition to the mean radius, we now also consider size spectra 
to address the collectional growth to larger particles.

The evolution of size spectra simulated with the Eulerian scheme with
3457 mass bins ($\beta=128$) is shown as blue lines in 
Figure~\ref{moments_shima_Nx32},
while the corresponding size spectra obtained with the swarm model
(collection scheme~II) with 32 particles
per grid point are shown as red curves.
The agreement between the Eulerian and Lagrangian schemes is good
at early times ($\tilde{t}\leq2000\,{\rm s}$), but at late times
($\tilde{t}=3000\,{\rm s}$) the size spectra from the Eulerian
approach is broader for the largest sizes ($r_{\max}=1000\,\mu{\rm m}$).
\citet{Shima09} found that the results of the super-droplet method
(collection scheme II)
agree fairly well with the numerical solution of a binned spectral method.
It is interesting to note that the size spectra simulated with the swarm model
(scheme~II) converge to those obtained with the Eulerian model with increasing
$N_p/N_{\rm grid}$. This can simply be explained by the fact that more
swarms contribute as potential collectional partners and thus ensure more
reliable statistics, which was also shown in the work of \citet{Shima09}.

\begin{figure}[t!]\begin{center}
\includegraphics[width=.8\textwidth]{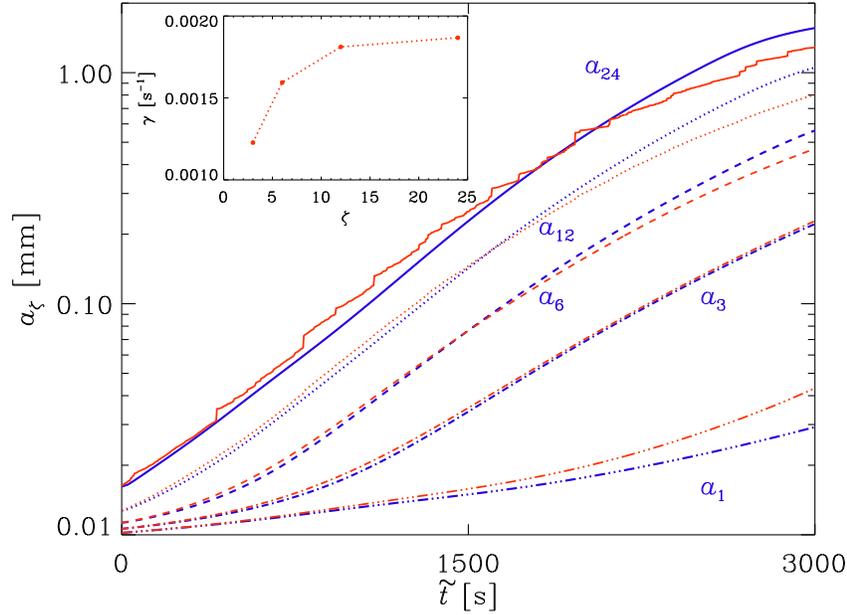}
\end{center}\caption[]{Same simulations as in 
Figure~\ref{moments_swarm_Euler}, but here we
only compare the scheme II and the Eulerian model.
Time evolution of $a_1$ (three-dotted dashed line), $a_3$ (dash-dotted lines),
$a_6$ (dash lines), $a_{12}$ (dotted lines), and $a_{24}$ (solid lines) 
of the size spectra are given. The inset shows the growth rates,
for several moments (fitted between $r$=$50\,\mu$m and $r$=$200\,\mu$m).
}\label{moments_shima_Euler}
\end{figure}

\subsection{Characterizing the size spectra}
\label{Characterizing}

\subsubsection{Estimating the extent of the size distribution}

The size spectra obtained in our simulations are rather broad and cover
nearly three orders of magnitude in radius (and six in mass).
Even at $\tilde{t}=3000\,{\rm s}$, the mean radius, $\overline{r}=a_1$,
has barely reached $30\,\mu{\rm m}$
(see Figure~\ref{moments_swarm_Euler}) and does not give any indication
about the width of the distribution.
The values of the higher moments $a_3$ and $a_6$ lie still only in the
middle of the size range; see Figure~\ref{moments_shima_Nx32}, where we
have marked the values of $a_1$, $a_3$, $a_6$, $a_{12}$,
and $a_{24}$ by arrows for $\tilde{t}=3000\,{\rm s}$.
We see that the value $a_{24}\approx1.5\,{\rm mm}$ characterizes
rather well the maximum radius of the distribution.
The actual maximum is at $\approx2.4\,{\rm mm}$,
but this value is rather noisy,
because it represents only a single data point in our simulated volume.
We have seen that our smallest and largest moments, $a_1$ and $a_{24}$,
conveniently bracket the extent of the size distribution.
However, for the type of size spectra presented here, the higher moments do not
contain any new or independent information, because all moments grow
exponentially at nearly the same rate; see Figure~\ref{moments_shima_Euler}.
This is quantified by the instantaneous growth rates,
$\gamma(\zeta)=d\ln(a_{\zeta})/dt$, which are found to be
around $0.0018{\rm s}^{-1}$ for large moments $\zeta$.
Thus, at least in the range $1000\,{\rm s}\leq\tilde{t}\leq3000\,{\rm s}$,
the value of $a_{24}$ is always approximately ten times larger than $a_{3}$.
However, this ratio can be different in different cases.
Knowing therefore the values of $a_1$ and $a_{24}$ gives a fairly reliable
indication about the full extent of the size spectra.

Given that the different moments are not independent, it should not be
too surprising that useful information can already be extracted from the
first three moments, which is at the heart of the so-called three-moment
bulk scheme \citep{Seifert16}.
This will be discussed next.

\subsubsection{Description in terms of the gamma distribution}

In the meteorological context, size spectra are often fitted to a
gamma distribution \citep{Berry74,Geoffroy10},
\begin{equation}
f(r)=nr^{\mu}e^{-\lambda r}\lambda^{\mu+1}/\Gamma(\mu+1).
\label{gamma_distribution}
\end{equation}
Here, the factor $\lambda^{\mu+1}/\Gamma(\mu+1)$, with $\Gamma$ being
the gamma function, is included so $f(r)$ is normalized such that
$\int_0^\infty f(r)\,{\rm d}r=n$.
In addition to $n$, it has $\mu$ and $\lambda$ as independent parameters,
which are all functions of time.
These are the basic parameters of the three-moment bulk scheme
\citep{Seifert16}.
An advantage of using the gamma distribution lies in the fact that
all moments can be calculated analytically.
Thus, one may ask for which values of $n$, $\mu$ and $\lambda$ do the moments
of the gamma distribution agree with those obtained here.
For given values of $a_1$ and $a_2$, we have
\begin{equation}
\mu=-\frac{(a_2/a_1)^2-2}{(a_2/a_1)^2-1}
=-\frac{a_2^2-2a_1^2}{a_2^2-a_1^2}
\quad\mbox{and}\quad
\lambda=\frac{\mu+1}{a_1}=\frac{a_1}{a_2^2-a_1^2}.
\label{shape}
\end{equation}
As stated above, the value of $n$ is given by the zeroth moment.
In Table~\ref{TamomText} we give the values of $n$ together with selected
normalized moments $a_\zeta$ as well as the resulting
parameters $\mu$ and $\lambda$.
We also give $\lambda^{-1}$, which has units of length and can be
compared with the normalized moments $a_\zeta$.

\begin{figure}[t!]\begin{center}
\includegraphics[width=.7\textwidth]{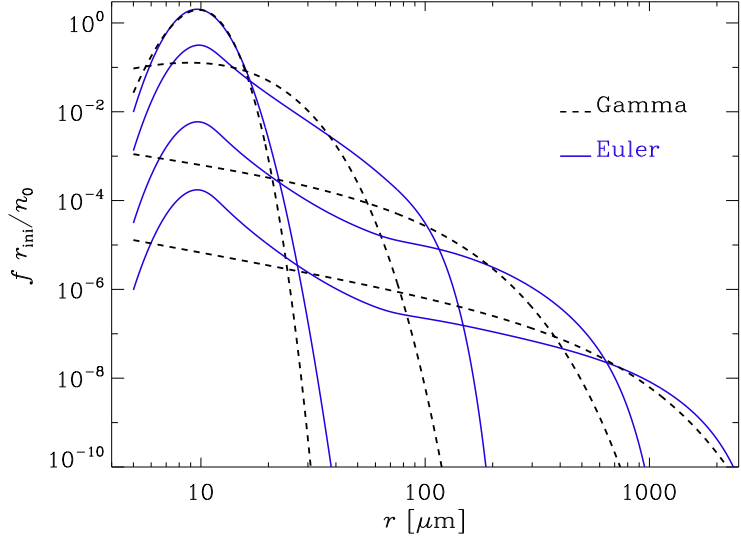}
\end{center}\caption[]{Same Eulerian simulation as in 
Figure~\ref{moments_swarm_Euler}.
The black lines show the gamma distribution estimated from the Eulerian 
simulation.
}\label{moments_gamma}
\end{figure}

Note that the value of $\mu$ decreases with time and approaches $-1$.
At the same time, $\lambda^{-1}$ increases and reaches $360\,\mu{\rm m}$
at the last time.
Although $\lambda^{-1}$ has units of length and tends to give an
indication about the cutoff of the distribution, its value 
is still five times smaller than that of $a_{24}$ and thus far away
from the maximum droplet radius.
This reinforces us in regarding $a_{24}$ as a useful measure of the
largest droplets.

\begin{table}[b!]\caption{
Zeroth moment $n$ (in units of ${\rm m}^{-3}$) together with selected
normalized moments $a_\zeta$ (in $\mu{\rm m}$), as well as the resulting
parameters $\mu$ (dimensionless), $\lambda$ (in $\mu{\rm m}^{-1}$),
and $\lambda^{-1}$ (in $\mu{\rm m}$) for Run~3C at different times.
Here, $\tilde{t}=1000\,t$.
}\vspace{12pt}\centerline{\begin{tabular}{ccrrrrrrrcc}
$t$ &  $n$    & $a_1$ & $a_2$ & $a_3$ & $a_6$&$a_{12}$&$a_{24}$
& $\mu\quad$ & $\lambda$ & $\lambda^{-1}$ \\
\hline
0&     $10^{11}$  &10.2& 10.4& 10.6& 11.3&  12.7&  16.2& 23.61 &2.412&0.4\\
1&$2\times10^{10}$&13.0& 15.0& 18.1& 32.4&  57.0&  87.6&  2.16 &0.242&4.1\\
2&$3.8\times10^8$ &17.7& 36.5& 68.3&168.0& 325.9& 530.2&$-0.69$&0.017&58 \\
3&$1.1\times10^7$ &29.3&106.7&221.3&562.4&1052.9&1560.7&$-0.92$&0.003&360\\
\label{TamomText}\end{tabular}}\end{table}

It turns out that the resulting profiles of the gamma distribution
capture the broad tail of the distribution remarkably.
This is shown in Figure~\ref{moments_gamma}, where we compare with
the actual size spectra.
It is obvious that the actual size distribution shows additional bumps,
notably at $\approx10\,\mu{\rm m}$.
This is completely missed when the parameters of the gamma distribution
are computed from the full data set.
Note also that we have employed a double-logarithmic representation
in Figure~\ref{moments_gamma}.
To model the bump at $10\,\mu{\rm m}$, one could again use the
gamma distribution, but now with moments that are based on $f(r)$
in a restricted size range, $r\leq30\,\mu{\rm m}$, for example.
This type of approach has been used by \cite{Seifert2001}, where they
describe the size spectra of cloud droplets and raindrops with
different distributions.

\cite{Seifert16} also found reasonable agreement between numerical
obtained size spectra and the corresponding gamma distribution.
They used the third and sixth moments of the distribution, but
in that case the corresponding expressions for $\mu$ and $\lambda$
are more complicated.
We show in Appendix~\ref{app:gamma} that the results do not
change much when using $a_3$ and $a_6$ to compute the parameters
of the gamma distribution.

Although the agreement between simulated size spectra and the gamma
distribution turns out to be reasonable, it only works for $\mu>-1$.
For smaller values of $\mu$, the function is no longer normalizable,
i.e., the integral over $f(r)\propto r^\mu$ diverges for $r\to0$
when $\mu+1<0$.
Most size spectra observed in meteorology have positive values of $\mu$
\citep{Seifert16}.
This is mainly because of the effect of evaporation, which is here neglected.
Evaporation would lead to a depletion of $f(r)$ for small values of $r$
and could lead to spectra that are more typical of a gamma distribution.
In our case, we have an approximate $r^{-1}$ fall-off over nearly two
orders of magnitude, followed by an exponential cutoff.
This is why the $\mu$ obtained from the moments turns out to be
so close to $-1$.
Thus, although the usefulness of the gamma distribution is
still being debated \citep{Khain15}, it can actually be remarkably good
provided one allows for small negative values of $\mu$.

\subsection{Inhomogeneous collection in a straining flow}
\label{Inho_collision}

Spatial variation in the flow leads to local concentrations and thus
to large peak values of $f(\mbox{\boldmath $x$},r,t)$ that shorten the collection time
$\tau_{\rm colle}$ \citep{1955_Saffman}.
Before studying the turbulent case, we consider first collectional growth
in a steady two-dimensional (2-D) divergence-free straining flow.
The straining flow is numerically inexpensive and easy to control 
and analyze compared with turbulence.

\begin{figure}[t!]\begin{center}
\includegraphics[width=\textwidth]{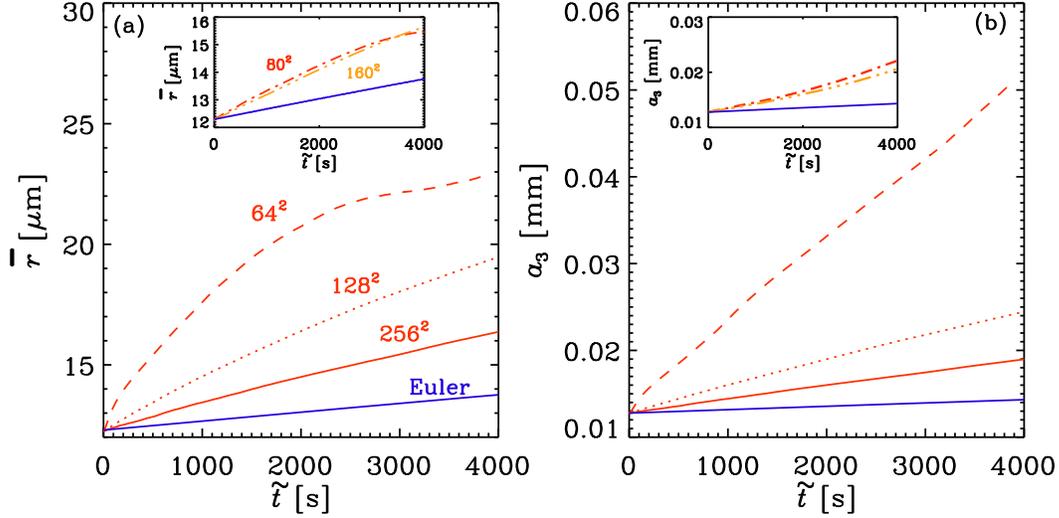}
\end{center}\caption[]{
Comparison of the evolution of (a) the mean particle size
and (b) $a_3$
in a straining flow for simulations
with the swarm approach at different grid resolutions.
Here, pure collection with
CIC particle interpolation algorithm has been used. 
The total number of swarms is $N_p=300,000$ while 
$D_p=0.05\,{\rm m^2}/{\rm s}$
and $\nu_p=0.01\,{\rm m^2}/{\rm s}$
are adopted in the Eulerian model.
The inset shows the case with NGP mapping instead of the CIC
first order interpolation for particle properties.
See Runs~7A, 7B, 7C, 7D, 7E, and 7F of
Table~\ref{runs} for simulation details.
}\label{pcomp_amean_L256condens53coag_forceStrain}\end{figure}

\subsubsection{Pure collection}
\label{pureCoa_strain}
We consider first the case of pure collection.
In Figure~\ref{pcomp_amean_L256condens53coag_forceStrain} we show the
time evolution of $\overline{r}$ for the swarm model with collection
scheme~II at different grid
resolutions ranging from $64^2$ to $256^2$ meshpoints.
Surprisingly, $\overline{r}$ grows more slowly as we increase
the mesh resolution of the swarm model.
Given that the swarm models seem to converge toward the Eulerian
model, we are confronted with the question of what causes the growth
of $\overline{r}$ in the swarm model to slow down at higher mesh resolution.
In this connection, we must emphasize that by default we use
the CIC algorithm to evaluate the gas properties
at the position of each Lagrangian particle.
As explained in section~\ref{imple}, the position of the shepherd particle
has no purpose other than to determine in which grid cell the swarm resides.
It is therefore {\emph not} better to use any kind
of interpolation in determining the value of the fluid variables at the
position of the swarm, but rather to use NGP mapping.
This will play an important role, as will be discussed now.
For the sake of solving equations~\eqref{dxidt} and
\eqref{dVidt}, the use of the CIC algorithm is perfectly valid,
but this would only be relevant for a direct Lagrangian tracking algorithm.
This can be understood by realizing that in the special case of particles
with vanishingly small inertia, the particles will follow their local
fluid cell, and hence, two particles will in the real world never collide.
However, if the CIC scheme is used for equations~\eqref{dxidt} and
\eqref{dVidt}, two swarms residing at different positions within the
{\em same} grid cell may have different velocities, and hence,
equation~\eqref{tauij1} may yield a collection.

Since the swarms are filling the entire
volume of the grid cell, this means that the two swarms will have different
velocities and exist in the same volume, and hence, the swarms may collide.
The larger grid cells yield potentially larger velocity differences 
between the particles, which explains why the collectional growth is 
larger for the coarser resolutions.
When NGP mapping is adopted, the artificial speedup disappears, as shown
in the inset of Figure~\ref{pcomp_amean_L256condens53coag_forceStrain}.

\begin{figure}[t!]\begin{center}
\includegraphics[width=\textwidth]{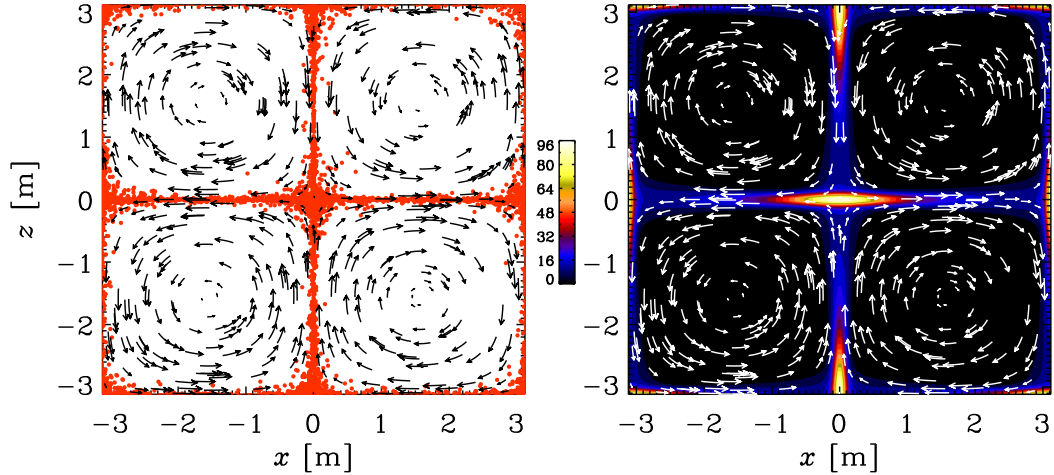}
\end{center}\caption[]{Visualization of flow
and particle field ($\tilde{t}$=1000 s) in a straining flow for simulations
with the swarm approach (left panel; red dotted curve in 
Figure.~\ref{pcomp_amean_L256condens53coag_forceStrain})
and Eulerian approach (right-hand panel; blue curve in 
Figure.~\ref{pcomp_amean_L256condens53coag_forceStrain}).
Here the radius of the particles is $r$=$128\,\mu{\rm m}$.
The swarms are represented by the red dots in the left-hand panel.
The contour map shows the spatial distribution of the number
density in the right panel.
The black and white arrows represent the velocity
vectors of the straining flow.
}\label{vf}
\end{figure}

However, the discrepancy between Lagrangian and Eulerian particle descriptions
is still strong for collectional growth in the straining flow as shown in
the inset of Figure~\ref{pcomp_amean_L256condens53coag_forceStrain}.
This is because that in a steady flow, the particles will end up
near the vertices of converging flow vectors and will therefore
be much more concentrated in the swarm model than what is possible
to represent in the Eulerian model.
This is evident by comparing the distribution of superparticles belonging
to a certain radius (here $128\,\mu{\rm m}$) with the corresponding distribution
function in the Eulerian model; see Figure~\ref{vf}.

\subsubsection{Combined condensation and collection}

\begin{figure}[t!]\begin{center}
\includegraphics[width=\textwidth]{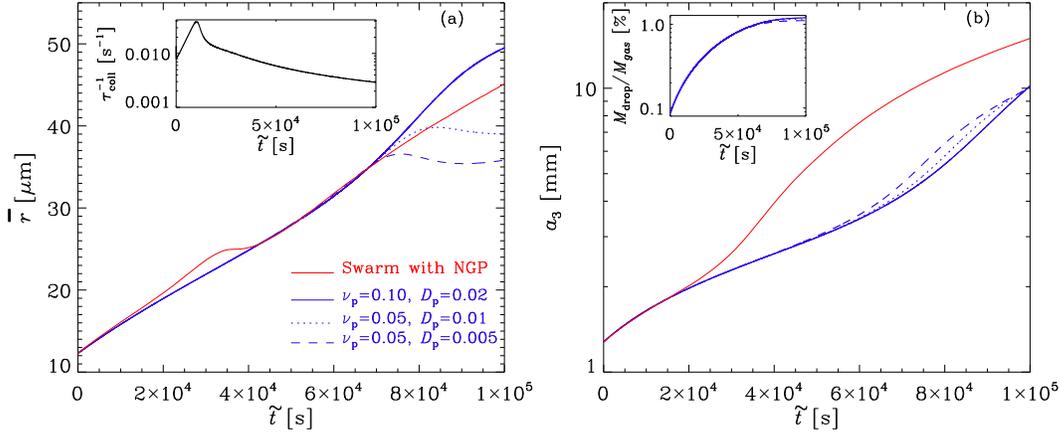}
\end{center}\caption[]{
Evolution of (a) $\overline{r}$ and (b) $a_3$ in the straining flow with combined
condensation and collection.
The different blue lines correspond to different amounts 
of artificial viscosity and enhanced Brownian diffusivity.
The inset of (a) shows the evolution of the inverse collection
timescale $\tau^{-1}_{\rm coll}$. The inset of (b) shows the evolution
of the mass ratio. The monotonic growth of the mass ratio demonstrates
that particles have not yet populated at the largest mass bin. 
The initial mean radius, supersaturation, and condensation parameter
is given by $r_{\rm ini}$=$12\,\mu{\rm m}$,
$s$=$0.01$, and $G=5\times10^{-11}\,{\rm m}^2/\,{\rm s}$, respectively,
and $k_{\max}$=$53$ with $\beta$=2.
See Runs~9A, 9B, 9C, 9D, and 9E of
Table~\ref{runs} for simulation details.
}\label{pcomp_L72condens22coag}\end{figure}

When both condensation and collection play a role, it is no longer
possible to define a unique timescale, and the solution depends on
both $\tau_{\rm cond}$ and $\tau_{\rm coll}$.
We consider here the straining flow using $r_{\rm ini}=12\,\mu{\rm m}$,
$G=5\times10^{-11}\,{\rm m}^2/\,{\rm s}$ and $s=0.01$,
which yields $\tau_{\rm cond}=144\,{\rm s}$.
We investigate the role that particle viscosity and Brownian diffusion
play in simulations using the Eulerian model.
The Brownian motion of the particles is usually small, so the
particle diffusion coefficient $D_p$ in equation~\eqref{eqn:dfdt} should be finite, but small.
Since it is assumed that the particle flows are relatively dilute, there
should be very little interaction between the different particle fluids,
except for the occasional collections. This implies that the
particle viscosity $\nu_p$ in equation~\eqref{particle_viscous} should be close to zero\footnote{Note that the
particle viscosity represents the coupling between the particle fluids --
not the drag coupling between the particles and the gas phase.}. 
For the Smoluchowski approach, both
$\nu_p$ and $D_p$ have to be made large in order to stabilize
the simulations in spatially extended cases.
It turns out that the values of these diffusion coefficients
have a surprisingly strong effect on the solutions, 
which is shown in Figure~\ref{pcomp_L72condens22coag}.
This could be due to the fact that the viscosity between the particle
fluids diffuses the momentum of the particles and thereby modifies
the collection rate.

Comparing now with the swarm approach, which avoids
artificial viscosity and enhanced Brownian diffusion
altogether, we see from Figure~\ref{pcomp_L72condens22coag}
that Eulerian and Lagrangian approaches agree with each other at
early times ($t<1000\,{\rm s}$).
After $1000\,{\rm s}$, both swarm and the Eulerian models follow
the same trend in the sense that the evolution of $\bar{r}$ shows a bump.
The bump occurs earlier for the swarm model than the Eulerian model.
In the extreme case that the artificial viscosity in the Eulerian model were
zero, the evolution of $\bar{r}$, as obtained from the swarm model, may come
closer that of the Eulerian model.
However, owing to the absence of a pressure term for particles,
discontinuities would develop in the Eulerian model that
destabilize the code if the viscosity and Brownian diffusion are too small.
Again, this may be a strong argument in favor of using the swarm model
for studying the collectional growth of cloud droplets. 

To relate the speed of evolution in Figure~\ref{pcomp_L72condens22coag}
to $\tilde{\tau}_{\rm coll}$, we plot in the inset of panel (a)
the inverse of its unscaled value, $\tau_{\rm coll}$,
as a function of time.
On average, we have $\tau_{\rm coll}\approx100$.
It is comparable to $\tau_{\rm cond}=144\,{\rm s}$ and both are
long compared with $\tau_{\rm cor}\approx1.4$.
The relevant quantity is the scaled value, $\tilde{\tau}_{\rm coll}$,
which is much larger $\approx10,000$.
This may suggest that the speed of growth is not governed by the
spatially averaged kernel, but by its value weighted toward regions
where the concentration is high.

We recall that growth of cloud droplets driven by pure collections
in the straining flow depends on the models (Eulerian and Lagrangian
models; see detailed comparisons in section~\ref{pureCoa_strain}).
This suggests that condensation has a ``regularizing'' effect
in that it makes
the overall evolution of $\overline{r}$ much less dependent on the
initial conditions and other model details.
This is due to the fact that the condensation process with constant 
positive supersaturation value leads to narrow size spectra of cloud 
droplets. 

Another interesting aspect is the bump 
in the evolution of the mean radius.  
At first glance it seems counterintuitive that $\overline{r}$ can actually
decrease during some time interval.
In Appendix~\ref{app:bump} we consider an example of four particles,
two large ones and two small ones.
If two small ones collide, we still have the two large ones, but only 3
particles in total after the collection, so the average radius increases
from 1/2 to 2/3.
On the other hand, if two large ones collide, we are still left with the
two small ones and one particle whose radius has only grown by a factor of
$2^{1/3}\approx1.26$ (the radius scales with the mass to the 1/3 power).
The average radius is therefore $2^{1/3}/3\approx0.42$, which
is less than the original mean radius, which is half the radius of the
large ones.

\subsection{Growth of droplets in 2-D turbulence}

Turbulence is generally believed to help bridging the size gaps in
both cloud droplet and planetesimal formation. In this section, 
pure turbulence-generated collections are simulated using 
both the Eulerian and Lagrangian models. 
We consider a 2-D squared domain of side length $L=0.5\,{\rm m}$
at a resolution of $512^2$ meshpoints,
with viscosity $\nu=5\times10^{-4}\,{\rm m}^2\,{\rm s}^{-1}$
(which is about 50 times the physical value for air),
average forcing wavenumber $k_{\rm f}\approx40\,{\rm m}^{-1}$,
i.e., $k_{\rm f} L/2\pi\approx3$,
and a root-mean-square velocity $u_{\rm rms}=0.8\,{\rm m}\,{\rm s}^{-1}$, resulting in
a Reynolds number of ${\rm Re}=u_{\rm rms}/\nu k_{\rm f}\approx40$.
Our choice of $k_{\rm f} L/2\pi\approx3$ corresponds to forcing at large
scales that are not yet too large to be affected by constraints
resulting form the Cartesian geometry.
The rate of energy dissipation per unit volume is
$\epsilon=2\nu{\langle {\mbox{\boldmath ${\sf S}$}^2}\rangle} \approx0.1\,{\rm m}^2\,{\rm s}^{-3}$ and the
turnover time is $\tau_{\rm to}=(u_{\rm rms}k_{\rm f})^{-1}\approx0.03\,{\rm s}$.
For the Lagrangian model, we use NGP mapping while for the Eulerian
model we adopt artificial viscosity and enhanced Brownian diffusivity for
the particles ($\nu_{\rm p}=D_{\rm p}=10^{-3}\,{\rm m}^2\,{\rm s}^{-1}$).

\begin{figure}[t!]\begin{center}
\includegraphics[width=\textwidth]{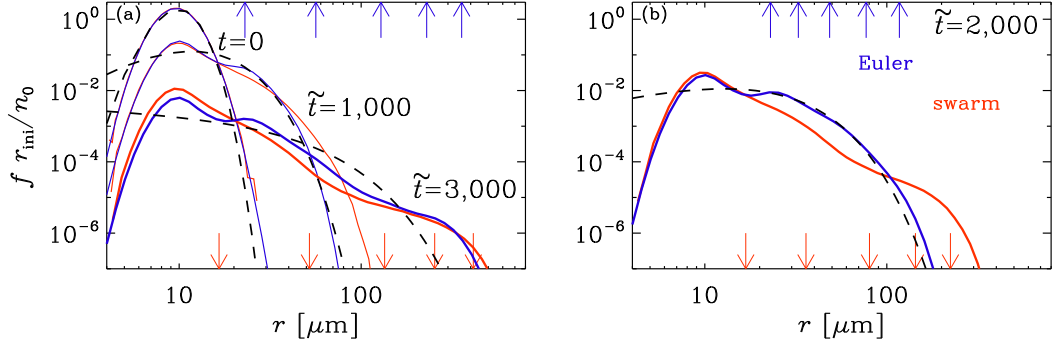}
\end{center}\caption[]{
Comparison of size spectra for Lagrangian (red lines) and Eulerian
(blue lines) approaches at different times in the presence of
2-D turbulence and no gravity nor condensation.
The arrows show the values of
different moments ($a_1$, $a_3$, $a_6$, $a_{12}$, and $a_{24}$, from left to right) of
the size spectra at time (a) $\tilde{t}=3000\,{\rm s}$
and (b) $\tilde{t}=2000\,{\rm s}$.
The largest departure between both approaches occurs for $\tilde{t}=2000\,{\rm s}$
and is plotted separately in the right-hand panel.
The black dashed curves are the fitted gamma distribution
of the Eulerian model.
See Runs~10A and 10B of Table~\ref{runs} for simulation details.
}\label{f_shima_Euler_con0_coa_grav0_tur_a010um_Re_2D}\end{figure}

\subsubsection{Size spectra}

Figure~\ref{f_shima_Euler_con0_coa_grav0_tur_a010um_Re_2D} 
shows the comparison of size spectra for the swarm and Eulerian models
in 2-D turbulence.
The agreement of the spectra as well as the high moments for both schemes
is good. 
Except for the latest time, the corresponding gamma distribution
agrees reasonably well with the simulated
size spectra, as shown by the black dashed curves in
Figure~\ref{f_shima_Euler_con0_coa_grav0_tur_a010um_Re_2D};
see selected moments and the parameters for the corresponding gamma distribution
in Table~\ref{TamomTurbText}.
Here we also give the parameters $\mu$ and $\lambda^{-1}$ of the
corresponding gamma distribution.
Again, $\mu$ becomes negative at the last time, but, unlike the
case with pure gravity (Table~\ref{TamomText}), its value is still
far away from $-1$.
Furthermore, $\lambda^{-1}$ is now 10 times smaller than $a_{24}$
and does therefore not represent the maximum droplet radius.

\begin{table}[htb]\caption{
Similar to Table~\ref{TamomText}, but now for Run~10A
with turbulence and no gravity.
Here, $\tilde{t}=100\,t$.
}\vspace{12pt}\centerline{\begin{tabular}{ccrrrrrrrcc}
$t$ &  $n$    & $a_1$ & $a_2$ & $a_3$ & $a_6$&$a_{12}$&$a_{24}$
& $\mu\quad$ & $\lambda$ & $\lambda^{-1}$ \\
\hline
 0&   $10^{10}$   &10.2&10.4&10.6& 11.3& 12.7& 16.2&$23.42$&2.394&0.4\\
10&$2.0\times10^9$&15.0&16.6&18.3& 23.3& 31.4& 43.1&$ 3.50$&0.300&3.3\\
20&$3.4\times10^8$&23.0&27.6&32.6& 48.1& 75.6&115.0&$ 1.25$&0.098&10\\
30&$6.7\times10^7$&22.8&34.7&56.4&132.2&243.2&388.0&$-0.23$&0.034&30\\
\label{TamomTurbText}\end{tabular}}\end{table}

We recall that good agreement is also observed in the case with gravity.
Hence, we conclude that the gamma distribution can reasonably well represent
the collectional size spectra.
This may provide a strong argument in favor of using the gamma distribution
when modelling cloud microphysics.

\subsubsection{Other numerical aspects}

It is worth noting that the MBR convergence of the Smoluchowski 
equation depends on the flow pattern. 
While gravitational collection is rather sensitive to MBR
(see Appendix~\ref{MBR_dependency_coag}),
it is much less sensitive for the straining flow and converges at
$k_{\rm max}\approx55$ in turbulence.

We emphasize that for the swarm model, the 
interpolation scheme of the tracked swarms does 
affect the results, but this does not seem to be the case for turbulence.
Turbulence continues to mix particles all the time while the straining
flow tends to sweep up particles into predetermined locations that do
not change.
We may therefore conclude that the restriction on the interpolation scheme
depends on the spatio-temporal properties of the flow.
Nevertheless, a high-order interpolation is not strictly applicable
to the swarm model.

It is worth noting that in the case with pure gravity,
the Eulerian model is rather sensitive to the presence or absence of the
${\cal M}_k$ term.
This is neither the case for turbulence nor for the straining flow as will
be discussed in Appendix~\ref{momentumC}.

\subsubsection{Comparison of computational cost}

Comparing the Lagrangian and Eulerian models in
Figure~\ref{f_shima_Euler_con0_coa_grav0_tur_a010um_Re_2D}, it is
worth noting that the Lagrangian one is clearly superior to the
Eulerian one in terms of CPU time for simulating the collectional
process in 2-D turbulence.
A similar conclusion was drawn by
\cite{Shima09}, who found the Lagrangian model to be computationally
faster than the Eulerian one.
We compare the computational cost between Eulerian and Lagrangian models
using the 2-D turbulence runs 10A and 10B (runs in 
Figure~\ref{f_shima_Euler_con0_coa_grav0_tur_a010um_Re_2D}), 
which have comparable accuracy;
see Table~\ref{runs} for details of these runs.
The Lagrangian model with $1.2\times10^6$ superparticles covers $217\,$s
in physical time within 24 hours of wall-clock time on 512 CPUs, while the
Eulerian model with 53 mass bins covers only $48\,$s in physical time
within 24 hours wall-clock time on 1024 CPUs.
This example demonstrates that the Lagrangian model is roughly
ten times more efficient than a comparable Eulerian one.

\subsubsection{Combined condensation and collection}

The combined condensational and collectional growth 
in turbulence is investigated as well. 
Again, the results are similar to the case with pure collectional growth 
due to the fact that the condensation process in the present study
with constant supersaturation is homogeneous.
In future studies, the supersaturation should be calculated self-consistently
and the effects of turbulence on the condensational growth
should be considered, similar to what was done previously
\citep{2014_Kumar,2015_Sardina}.

\section{Conclusion}

The combined collectional and condensational growth 
of cloud droplets is studied in numerical simulations where the
gas phase is solved on a mesh, while the particle phase is approximated
by a point particle approach and is treated
either by an Eulerian or a Lagrangian formalism. 
It is found that the Lagrangian approach agrees well with
the analytic solution of condensational growth.
By contrast, the Eulerian approach requires high resolution in the
number of mass bins to avoid artificial speedup of the growth rate,
which agrees with previous findings \citep{1990_Ohtsuki,Dullemond_2014}. 
It is worth noting that the MBR dependency is closely related to 
the temporal and spatial properties of the flow.
The dependency is strongest for gravity [\mbox{\boldmath
$u=0$} in equation~\eqref{turb}], less strongly for the
straining flow, and weak for turbulence.

A detailed comparison of the collectional size spectra between the
Lagrangian and Eulerian models demonstrates consistency between the two, 
especially when both condensation and collection are included.
This suggests that condensation has a regularizing effect and makes
the overall evolution of the mean radius less dependent on details
such as the precise form of the initial condition or discretization
errors that might affect the early evolution.
However, the evolution of the mean radius, i.e., the ratio of the two
lowest (first and zeroth) moments of the size distribution function,
is a rather insensitive measure of particle growth.
This is also seen in the fact that the mean particle radius often
increases by not much more than a factor of three, while the size
distribution can become rather broad and even millimeter-sized particles
can be produced within a relatively short time.
The mean particle radius is also not the most relevant diagnostics in that it
does not characterize properly the growth of the largest particles.
In fact, as we have shown in Appendix~\ref{app:bump}, the mean radius actually
{\em decreases} when two large particles collide.
This is somewhat counterintuitive, but actually quite natural.
When two very small particles collide, the sum of all radii does
basically not change, but the number of particles decreased by one,
so the average increases.
By contrast, when two large particles collide, the particle number
again decreases by one, but the sum of the radii decreases from 2 to
$2^{1/3}\approx1.26$, so the average also decreases.

Remarkably, we found that the simulated tails of the size spectra
agree fairly well with those obtained from the gamma distribution
for negative $\mu$. More importantly, the agreement is good
both for gravity and turbulence cases.

When studying pure collection, the Eulerian approach yields
satisfactory results only when the mass bins are sufficiently fine.
Furthermore, for collections in the case of a straining flow, it is
found that the Eulerian approach requires artificially large viscosity
and Brownian diffusivity
for keeping the resulting shocks in the particle fluid resolved.
Because of this, it seems that for future studies of the effect of
turbulence on condensational and collectional growth of particles,
the Lagrangian swarm approach would be most suitable.
However, several precautions have to be taken.
First, the symmetric collection scheme~II \citep{Shima09}
is to be preferred because it shows less scatter in the mean radius
than the asymmetric scheme~I.
This is because in scheme~I the particle number is adjusted to keep
the total mass in the swarm constant.
Second, when interpolation of the gas properties at the position of
each Lagrangian particle is invoked (for example the CIC algorithm
or the triangular shaped cloud scheme), both collection schemes yield
artificially increased collection rates. This is because two swarms within
the same grid cell may always collide since the interpolation of the 
fluid velocity results in a velocity difference between the two swarms.
This causes a speedup of the collection rate already at early times.
At higher grid resolution, the interpolated velocity differences are
smaller, which reduces the collectional growth.
Therefore, it is best to map the gas properties to just the nearest grid
point, which is found to yield converged results even at low resolution.

A shortcoming of the Eulerian model is that self-collections are strictly
impossible.
This should be mitigated by using finer mass bins, but it
turns out that finer mass bins do not change the collection rate
at early times, but rather decrease it at later times.
This indicates that the contribution of self-collections to the 
collection rate is relatively small; see Appendix~\ref{SelfCollision}
for details.

The discrepancy between Lagrangian and Eulerian particle descriptions
is particularly strong in the time-independent straining flow.
This is because particles tend to be swept into extremely narrow lanes,
which leads to high concentrations that can never be achieved with the
Eulerian approach, in which sharp gradients must be smeared out by
invoking artificial viscosity and large Brownian diffusivity.
On the other hand, we are here primarily interested in turbulent flows
that are always time-dependent, which limits the amount of particle
concentration that can be achieved in a given time.
In that case, the discrepancies between Eulerian and Lagrangian approaches
are smaller at early times, but there are still differences in the
evolution of the mean radius at late times.
This can easily be caused by changes in the relative importance of
collections of large and small particles.
This is confirmed by the fact that the size distribution spectra in the
turbulent case are more similar for Lagrangian and Eulerian approaches 
than in the straining flow. 

Our present work neglects local and temporal changes in the
supersaturation. 
In future studies, we will take into account that the supersaturation
increases (decreases) as a fluid parcel 
rises (falls) and that droplet condensation (evaporation) act as sinks (sources)
of supersaturation.
We would then be able to account for the fact that the total water content
should remain constant and that the supersaturation would become
progressively more limited as water droplets grow by condensation.
Another important shortcoming is our assumption of perfect collection
efficiency, which resulted in artificially rapid cloud droplets growth.
Alleviating these restrictions will be important tasks for future work.
Furthermore, we have here only considered 2-D turbulence.
Extending our work to 3-D is straightforward, but our conclusions
regarding the comparison of different schemes should carry over to 3-D.

\acknowledgments

We thank Nathan Kleeorin, Dhrubaditya Mitra, and Igor Rogachevskii for
useful discussions.
We also thank the anonymous referees for constructive comments
and suggestions that lead to substantial improvements in the manuscript.
This work was supported through the FRINATEK grant 231444 under the
Research Council of Norway, the Swedish Research Council grant 2012-5797,
and the grant ``Bottlenecks for particle growth in turbulent aerosols''
from the Knut and Alice Wallenberg Foundation, Dnr.\ KAW 2014.0048.
The simulations were performed using resources provided by
the Swedish National Infrastructure for Computing (SNIC)
at the Royal Institute of Technology in Stockholm.
This work utilized the Janus supercomputer, which is supported by the
National Science Foundation (award number CNS-0821794), the University
of Colorado Boulder, the University of Colorado Denver, and the National
Center for Atmospheric Research. The Janus supercomputer is operated by
the University of Colorado Boulder.
The source code used for the simulations of this study, the {\sc Pencil Code},
is freely available on \url{https://github.com/pencil-code/}.
The input files as well as some of the output files of the simulations
listed in Table~\ref{runs} are available under
\url{http://www.nordita.org/~brandenb/projects/SwarmSmolu_numerics/}.

\begin{appendices}
\numberwithin{figure}{section}
\numberwithin{equation}{section}
\numberwithin{table}{section}

\section{Upwinding scheme for a nonuniform mesh}
\label{AppA}

In the presence of condensation alone, the evolution equation for
$f(r,t)$ as a function of radius $r$ and time $t$ is given by
\begin{equation}
{\partial f\over\partial t}=-{\partial\over\partial r}(f C),
\label{radius_advection}
\end{equation}
where $C\equiv{\rm d} r/{\rm d} t$ and is given by equation~\eqref{cond_eq}.
Thus, we have
\begin{equation}
{\partial f\over\partial t}=-A{\partial\over\partial r}\left({f\over r}\right)
\label{radius_advection_novera}
\end{equation}
where $A=Gs$ is assumed independent of $r$; see equation~(13.14) of \citet{Sei06}.
It can be seen from the form of the analytic solution that there will
be a discontinuity at $r^2=2At$, which is numerically difficult to handle.
In particular, it is difficult to ensure the positivity of $f$.
For these reasons, a low-order upwind scheme is advantageous.
Furthermore, expanding the RHS of equation~\eqref{radius_advection_novera}
using the quotient rule,
\begin{equation}
\frac{\partial f}{\partial t}=\frac{A}{r^2}f
-\frac{A}{r}\frac{\partial f}{\partial r},
\label{danger}
\end{equation}
it is obvious that the first term in isolation would lead to exponential growth
of $f$ proportional to $\exp(At/r^2)$, which must be partially canceled
by the second term.
If the cancellation is numerically imperfect, $f(r,t)$ will indeed
grow exponentially, which tends to occur in regions where $r^2<2At$,
i.e., where $f$ should vanish.
For nonuniform mesh spacing, $r_k$ with $k=1$, 2, ..., $k_{\max}$, the
first-order upwind scheme can be written as
\begin{eqnarray}
{\partial f_k\over\partial t}=c_k^+ {f_{k+1}\over r_{k+1}}
                             +c_k^0 {f_{k  }\over r_{k  }}
                             +c_k^- {f_{k-1}\over r_{k-1}}
\label{formula}
\end{eqnarray}
with
\begin{equation}
c_k^\pm=\pm{\textstyle\frac{1}{2}}{|A|\mp A\over r_{k\pm1}-r_{k}},\quad
c_k^0=-c_k^+ -c_k^-.
\label{both2}
\end{equation}
On the boundaries of the radius bins at $k=1$ and $k_{\max}$,
equation~\eqref{formula} cannot be used unless we make an assumption
about the nonexisting points outside the interval $1\leq k\leq k_{\max}$.
For example, for $k=k_{\max}$, the coefficient $c_k^+$ would multiply
$f_{k+1}/r_{k+1}$, which is not defined.
Therefore, a simple assumption is to set $c_k^+=0$.
However, $c_k^+$ also enters in the expression for $c_k^0$, which
is the factor in front of $f_{k}/a_{k}$.
The coefficient $c_k^+$ can only be nonvanishing when $A<0$.
If we were to omit $c_k^+$ in the expression for $c_k^0$,
then, for $A<0$, the value of $f_k$ would not evolve at $k=k_{\max}$
and would be frozen.
Thus, the non-existing points lead to an unphysical situation.
It would be natural to assume that at $k=k_{\max}$, $f_k$ should decay
with time at a rate $-(|A|-A)/r_{k}$.
Therefore, assume
\begin{equation}
c_k^+=0,\quad
c_k^0=-(|A|-A)/r_{k}-c_k^-
\quad\mbox{(for $k=k_{\max}$)}
\label{both3}
\end{equation}
and $c_k^-$ unchanged, and analogously
\begin{equation}
c_k^-=0,\quad
c_k^0=-(|A|+A)/r_{k}-c_k^+
\quad\mbox{(for $k=0$)}
\end{equation}
and $c_k^+$ unchanged.

\section{Momentum conservation solution of the Eulerian model}
\label{momentumC}

The purpose of this appendix is to derive the momentum-conserving
velocity kick $\bm{{\cal M}}_k$ in equation~(\ref{dvdt})
and to demonstrate how it works.
Each collection event involves three partners, which we denote
by subscripts $i$, $j$, and $k$, where $k$ is the result of
the collection between $i$ and $j$.
Mass conservation implies that $f_im_i+f_jm_j+f_km_k$ is constant,
i.e., its time derivative vanishes.
Likewise, momentum conservation implies that
\begin{equation}
\frac{\partial}{\partial t}\left(
f_im_i\bm{v}_i+f_jm_j\bm{v}_j+f_km_k\bm{v}_k\right)=0.
\end{equation}
The time derivatives of $f$ caused by collections is ${\cal T}$,
while that of $\bm{v}$ is $\bm{{\cal M}}$.
However, only the resulting particle $k$ will suffer a kick,
while $i$ and $j$ do not, so we have
\begin{equation}
{\cal T}_i m_i \bm{v}_i + {\cal T}_j m_j \bm{v}_j
+ {\cal T}_k m_k \bm{v}_k + f_k m_k \bm{{\cal M}}_k=0.
\label{momcons1}
\end{equation}
As seen from equation~\eqref{Tcoag2}, for the collection of
$i$ and $j$, the increase in $f_k$ is given by
\begin{equation}
{\cal T}_k = K_{ij} f_i f_j \, \frac{m_i+m_j}{m_k},
\label{Ttermk}
\end{equation}
while the corresponding decreases in both $f_i$ and $f_j$ are
\begin{equation}
{\cal T}_i = {\cal T}_j = -K_{ij} f_i f_j,
\label{Ttermsij}
\end{equation}
which evidently obeys mass conservation, i.e.,
${\cal T}_i m_i + {\cal T}_j m_j + {\cal T}_k m_k = 0$.
Inserting equations~(\ref{Ttermk}) and (\ref{Ttermsij}) into
equation~(\ref{momcons1}) and solving for $\bm{{\cal M}}_k$ yields
\begin{equation}
\bm{{\cal M}}_k=\frac{1}{f_k m_k} K_{ij} f_i f_j
\left[ m_i \bm{v}_i + m_j \bm{v}_j - (m_i + m_j) \bm{v}_k \right].
\label{momcons2}
\end{equation}

\begin{table}[t!]
\caption{Total particle momentum in ${\rm kg}{\rm m}^{-2}{\rm s}^{-1}$
after three different times using the Eulerian model.}
\centering
\begin{tabular}{crrrrrrr}
\hline
case & ${\cal T}$ & ${\cal M}$ & $g$ & $t=0.0\,$s & $t=0.1\,$s & $t=1\,$s & $t=10\,$s \\
\\
\hline
A & $    0$ & $    0$ & $    0$ & $ 0.8042$ & $ 0.8042$ & $ 0.8042$ & $  0.8042$  \\
B & $\neq0$ & $    0$ & $    0$ & -- & $ 0.3386$ & $ 0.0012$ & $  0.00$  \\
C & $\neq0$ & $\neq0$ & $    0$ & -- & $ 0.8035$ & $ 0.8032$ & $  0.75$  \\
D & $    0$ & $    0$ & $\neq0$ & -- & $ 0.3070$ & $-4.1679$ & $-48.92$  \\
E & $\neq0$ & $    0$ & $\neq0$ & -- & $-0.1586$ & $-4.9709$ & $-49.72$  \\
F & $\neq0$ & $\neq0$ & $\neq0$ & -- & $ 0.3063$ & $-4.1673$ & $-45.51$  \\
\hline
\multicolumn{7}{p{0.6\textwidth}}{The initial parameters are:
$v_1$=$1\,\rm{m}\,\rm{s}^{-1}$ and $v_2$=$2\,\rm{m}\,\rm{s}^{-1}$ at
radius bins $r_1$=$100\,\mu{\rm m}$ and $r_2$=$112\,\mu{\rm m}$
($\times2^{1/6}$ larger) with $n_0$=$10^8 {\rm\, m^{-3}}$ distributed
evenly over the first two mass bins.
}\end{tabular}
\label{MC}
\end{table}

We give in Table~\ref{MC} the values of the total momentum of all
particles in the Eulerian model, $\sum \hat{f}_i m_i \bm{v}_i$,
at three different times for a model without spatial extent (0-D).
Initially, we have two mass bins with velocities 1 and 2, which
leads to collectional growth if ${\cal T}\neq0$.
Drag with the gas is here neglected.
In the absence of gravity, the total momentum is the same for all
three times when there is no collection (${\cal T}=0$).
For ${\cal T}\neq0$, there is a dramatic change of momentum
if the ${\cal M}$ term is neglected (case~B).
With the ${\cal M}$ term included, momentum is reasonably well conserved
(compare case~C with case~A).
In the presence of gravity, the momentum changes just because of
gravitational acceleration (cases~D--F).
However, we would not expect the total momentum to change dramatically
when we allow for collection (${\cal T}\neq0$).
We see that without the ${\cal M}$ term the total momentum departs
significantly from the case without collection (case~E), while
with the ${\cal M}$ term included, the values of total momentum
are similar to those without collection (compare case~F with case~D).
This validates the implementation of the momentum conserving term.

\begin{figure}[t!]\begin{center}
\includegraphics[width=\textwidth]{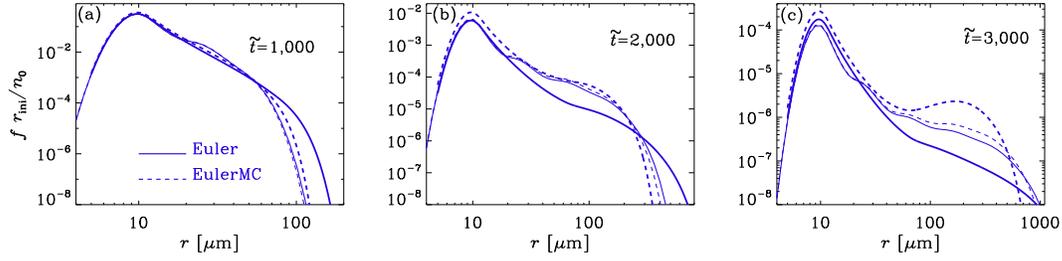}
\end{center}\caption[]{Same as Figure~\ref{moments_shima_Nx32},
but with the Eulerian model with momentum conservation
(blue dashed lines, denoted by ``EulerMC'') included.
Here we only compare EulerMC and Euler.
Thick lines: $\beta=8$; thin lines: $\beta=2$.
See additional Runs 3A, 3D, and 3E in Table~\ref{runs_appendix}
for simulation details.
}
\label{moments_EulerMC_comp}\end{figure}

Let us now discuss the effect of the momentum conserving correction in the
context of gravitational collection.
This is shown in Figure~\ref{moments_EulerMC_comp}, where we compare
size spectra for $\beta=2$ and $8$ with and without the ${\cal M}$ term.
It turns out that without the ${\cal M}$ term, the growth of large
droplets is increased when the MBR is large ($\beta=8$).
This is not the case, however, when the ${\cal M}$ term is included,
which leads to a much slower growth of the largest droplets.
On the other hand, as demonstrated above, the ${\cal M}$ term leads to a
decrease of the momentum of the large droplets, which explains the
absence of particles above $1\,{\rm mm}$ at $\tilde{t}=3000\,{\rm s}$
and the increase at smaller radii.

\begin{figure}[t!]\begin{center}
\includegraphics[width=\textwidth]{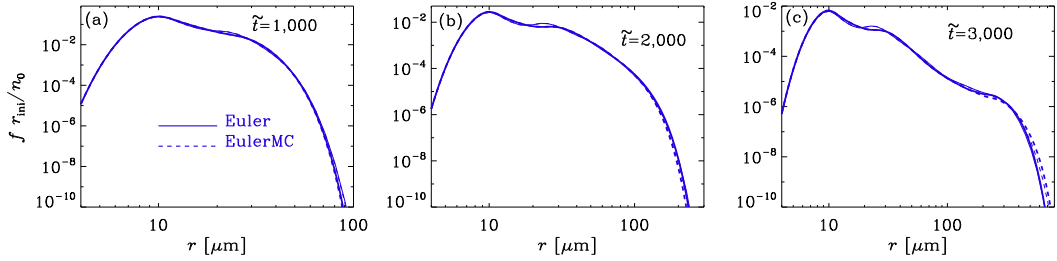}
\end{center}\caption[]{The effect of the momentum
conserving term for a turbulent flow
(dashed lines, denoted by ``EulerMC'') compared with
the case without it (denoted by ``Euler''), same as in 
Figure~\ref{f_shima_Euler_con0_coa_grav0_tur_a010um_Re_2D}.
Thick lines: $\beta=8$; thin lines: $\beta=2$.
See additional Runs 10C, 10D, and 10E in Table~\ref{runs_appendix}
for simulation details.
}\label{f_comp_EulerMC_tur}\end{figure}

Remarkably, in turbulence, the evolution of the
size spectra are almost the same with or without
momentum correction term.
This is shown in Figure~\ref{f_comp_EulerMC_tur}.
It is still unclear why the effect of the momentum
correction term depends so strongly on the flow pattern.
Further investigation is required to understand this
in the future work.
However, one might speculate that the
momentum conservation correction accumulates numerical errors with
increasing number of mass bins, so it is unclear that this procedure
leads to more accurate results.

\begin{table}[t!]
\caption{Summary of additional simulations presented in the appendix.}
\centering
\setlength{\tabcolsep}{3pt}
\begin{tabular}{l c c c c c c c c c c c p{1.5cm}}
\hline
Run & Scheme & Dim & $L$ (m) & $N_p$ & $N_{\rm grid}$ & IM & Processes &
$\beta$ & $n_0$ (${\rm m^{-3}}$) & Flow & $D_p$ (${\rm m^2}/{\rm s}$) & $\nu_p$ (${\rm m^2}/{\rm s}$) \\
\hline
1B &Eu & 0-D & $0.5$ & -- & -- & -- & Con & $2$ & $10^{8}$ & -- \\
2A &Eu & 0-D & $0.5$ & -- & -- & -- & Con & $8$ & $10^{8}$ & -- \\
3A &Eu & 0-D & $0.5$ & -- & -- & -- & Col & $2$ & $10^{8}$ & grav \\
3B &Eu & 0-D & $0.5$ & -- & -- & -- & Col & $32$ & $10^{8}$ & grav \\
3D &EuMC& 0-D & $0.5$ & -- & -- & -- & Col & $2$ & $10^{10}$ & grav \\
3E &EuMC& 0-D & $0.5$ & -- & -- & -- & Col & $128$ & $10^{11}$ & grav \\
4A &SwI & 3-D & $0.5$ & $2N_{\rm grid}$ &  $32^3$ &CIC& Col & -- & $10^{10}$ & grav \\ 
4C &SwI & 3-D & $0.5$ & $8N_{\rm grid}$ &  $32^3$ &CIC& Col & -- & $10^{10}$ & grav \\ 
4D &SwII & 3-D & $0.5$ & $2N_{\rm grid}$ &  $32^3$ &CIC& Col & -- & $10^{10}$ & grav \\ 
4E &SwII & 3-D & $0.5$ & $4N_{\rm grid}$ &  $32^3$ &CIC& Col & -- & $10^{10}$ & grav \\ 
4F &SwII & 3-D & $0.5$ & $8N_{\rm grid}$ &  $32^3$ &CIC& Col & -- & $10^{10}$ & grav \\ 
4G &SwI & 3-D & $0.5$ & $4N_{\rm grid}$ &  $32^3$ &CIC& Col & -- & $10^{10}$ & grav \\ 
5A &SwII & 3-D & $0.5$ & $2\times10^3$ &  $N_p/4$ &CIC& Col & -- & $10^{10}$ & grav \\ 
5B &SwII & 3-D & $0.5$ & $16\times10^3$ &  $N_p/4$ &CIC& Col & -- & $10^{10}$ & grav \\ 
5C &SwII & 3-D & $0.5$ & $442\times10^3$ &  $N_p/4$ &CIC& Col & -- & $10^{10}$ & grav \\ 
5D &SwII & 3-D & $0.5$ & $1024\times10^3$ &  $N_p/4$ &CIC& Col & -- & $10^{10}$ & grav \\ 
8A &Eu & 2-D & $2\pi$ & -- & $80^2$ &--& Col & $2$ & $10^{10}$ & strain & $0.01$ & $0.05$ \\ 
8B &Eu & 2-D & $2\pi$ & -- & $80^2$ &--& Sym & $2$ & $10^{10}$ & strain & $0.01$ & $0.05$ \\ 
8C &Eu & 2-D & $2\pi$ & -- & $80^2$ &--& Ave & $2$ & $10^{10}$ & strain & $0.01$ & $0.05$ \\ 
8D &Eu & 2-D & $2\pi$ & -- & $80^2$ &--& Sym & $4$ & $10^{10}$ & strain & $0.01$ & $0.05$ \\ 
8E &SwII & 2-D & $2\pi$ & $5\times10^4$ &  $80^2$ &NGP& Col & -- & $10^{10}$ & strain \\ 
8F &Eu & 2-D & $2\pi$ & -- & $80^2$ &--& Col & $4$ & $10^{10}$ & strain & $0.01$ & $0.05$ \\ 
8G &Eu & 2-D & $2\pi$ & -- & $128^2$ &--& Both & $4$ & $10^{8}$ & strain & $0.01$ & $0.05$ \\ 
10C &EuMC & 2-D & $0.5$ & -- & $512^2$ &--& Col  & $2$ & $10^{10}$ & turb & $0.001$ & $0.001$ \\ 
10D &EuMC & 2-D & $0.5$ & -- & $512^2$ &--& Col  & $4$ & $10^{10}$ & turb & $0.001$ & $0.001$ \\ 
10E &Eu & 2-D & $0.5$ & -- & $512^2$ &--& Col  & $4$ & $10^{10}$ & turb & $0.001$ & $0.001$ \\ 
\hline
\multicolumn{13}{p{\textwidth}}{Here, the abbreviations are the
same as the ones in Table~\ref{runs} but with additional
abbreviations listed below. ``Sym'' refers to collection with
symmetric self-collection invoked in Eulerian model, ``Ave''
refers to collection with average self-collection invoked in
Eulerian model, ``EuMC'' refers to the Eulerian model with
momentum conservation invoked.}
\end{tabular}
\label{runs_appendix}
\end{table}

\section{MBR dependency for condensation}
\label{MBR_dependency}

\begin{figure}[t!]\begin{center} 
\includegraphics[width=\textwidth]{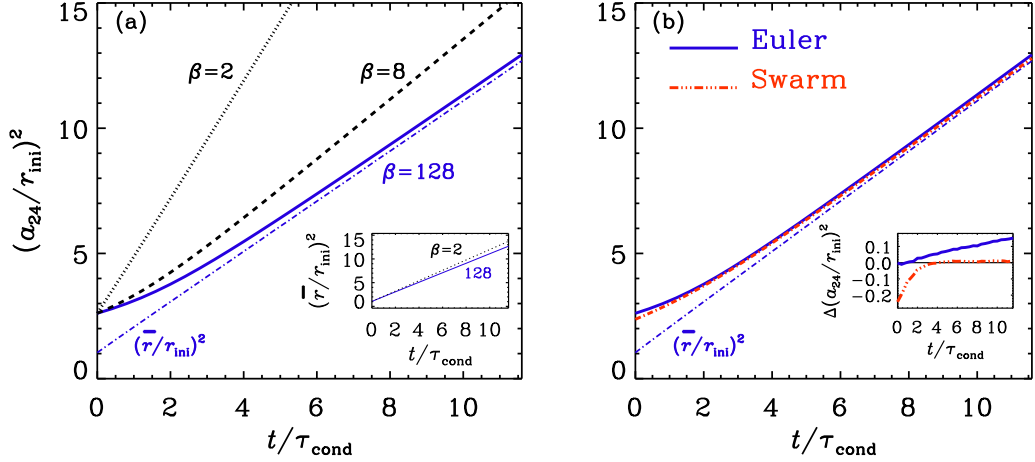}
\end{center}\caption[]{
Same run as in Figure~\ref{psol_comp_5em6}, but with different $\beta$
for the Eulerian model.
Comparison of the time evolution [in units of $\tau_{\rm cond}$,
as given by equation~\eqref{eqn:tau}]
of $(a_{24}/r_{\rm ini})^2$ for $\beta$=2 (dotted red), 8 (dashed black),
and 128 (solid blue) for condensation, together with $(\overline{r}/r_{\rm ini})^2$
for $\beta$=128 (dash-dotted blue) using the mass bin interval $2$--$20\,\mu{\rm m}$.
The inset shows $(\bar{r}/r_{\rm ini})^2$ for $\beta$=2 (dotted red)
and 128 (solid black).
The right panel shows a comparison of
the 24th moment between the Eulerian model with 1281 mass bins (solid blue)
and the swarm model with collection scheme~I,
$N_p$=$10,000$, and $N_{\rm grid}$=$16^3$ (red).
The inset shows the difference of the squares to the analytic solution
for the Eulerian model (solid blue) and the swarm model (triple-dot-dashed red).
Here, the parameters for condensation and the initial conditions
are the same as for Figure~\ref{psol_comp_5em6}.
See additional Runs~1B and 2A of Table~\ref{runs_appendix}
for simulation details.}
\label{pmoms_comp}
\end{figure}

As discussed in section~\ref{CondensationExperiments},
one needs extremely large MBR to model condensation accurately.
This becomes particularly critical when using 
logarithmic spacing on a mesh with small values of $\delta$.
The purpose of this appendix is to demonstrate the effects on
the tails of the distribution.
In Figure~\ref{pmoms_comp} we show $(a_{24}/r_{\rm ini})^2$ for different 
MBR ($\beta$) and compare with $(\overline{r}/r_{\rm ini})^2$.
In the Eulerian model, we consider the values $\beta=2$, 8, and 128
over the mass bin interval $2$--$20\,\mu{\rm m}$, so the number of bins are
$k_{\max}-1=20$, 80, and 1280, respectively.
In the Lagrangian model, we use $N_p=10,000$ and $N_{\rm grid}=16^3$,
so $N_p/N_{\rm grid}\approx2.4$.
At higher MBR, the $a_\zeta$ for different values
of $\zeta$ converge to the same value, but not at low
MBR (see the inset of the first panel).
This can have a lasting effect on the growth of the higher moments in the
sense that the slope in Figure~\ref{pmoms_comp} is increased at all later times.
This is consistent with earlier findings \citep{1990_Ohtsuki,Dullemond_2014}.
When collection is included, the artificially broadened tails in the
distribution can be particularly dangerous, because they would have
a strong effect on the rate of collection, which would be faster
when the $a_\zeta$ for large values of $\zeta$ are increased
by the artificially broadened size distribution.
In the right hand-panel of Figure~\ref{pmoms_comp}, we compare
$(a_{24}/r_{\rm ini})^2$
for both the swarm model and the high MBR Eulerian
simulation. 
The inset shows for both the swarm and the Eulerian models the departure,
\begin{equation}
\Delta [(a_{24}/r_{\rm ini})^2]=(a_{24}/r_{\rm ini})^2-(a_{24}/r_{\rm ini})^2_{\rm analyt},
\end{equation}
from the analytic solution.
We see that at late times the swarm model agrees perfectly, while the
Eulerian one shows small but persistent departures, as mentioned before.
From this it is clear that the swarm model reproduces the high MBR
Eulerian simulation rather accurately, but at a much lower computational cost.

\begin{figure}[t!]\begin{center}
\includegraphics[width=\textwidth]{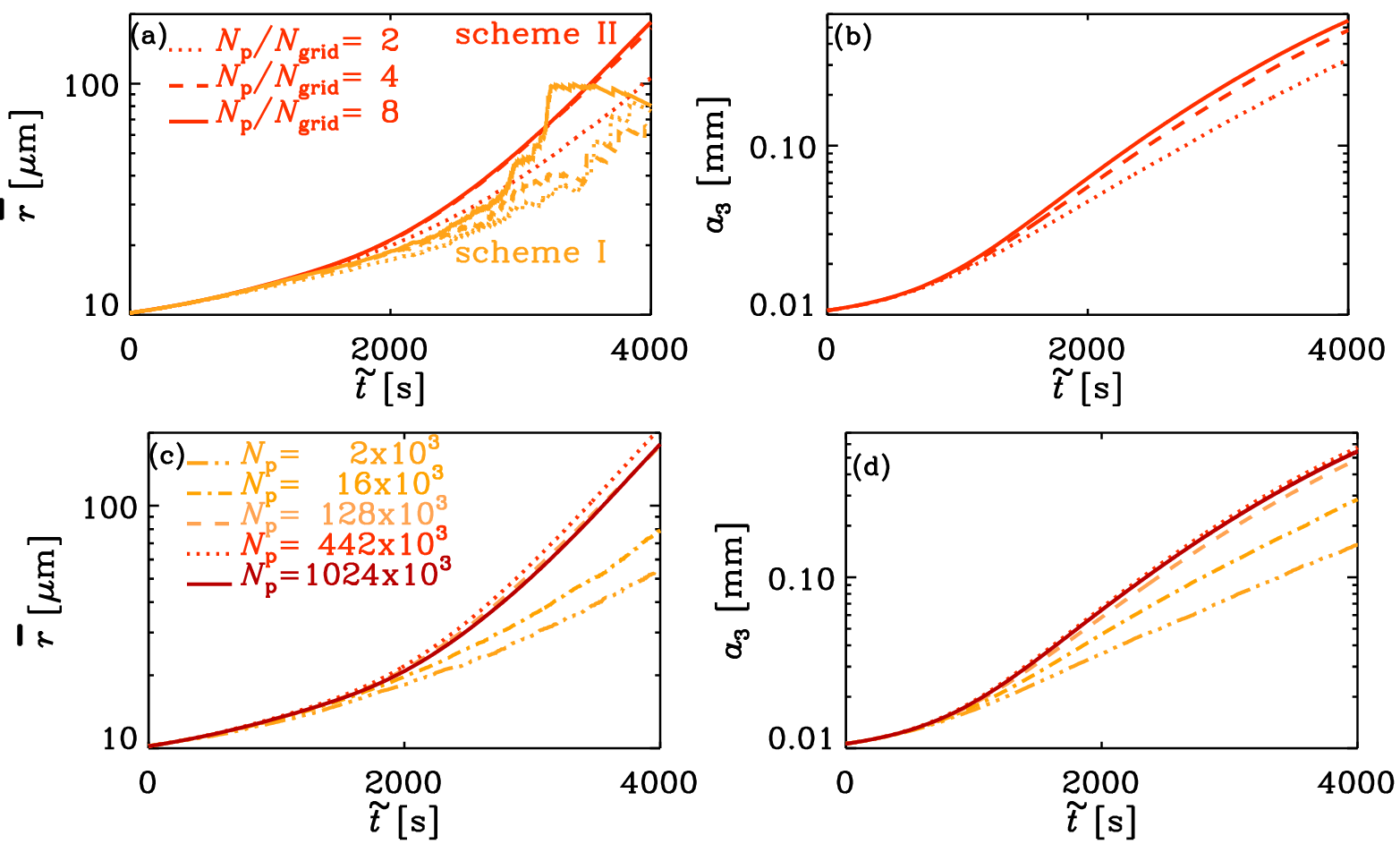}
\end{center}\caption[]{
Same as Figure~\ref{moments_swarm_Euler}, but here we only study the
statistical convergence properties of swarm model.
Upper panels: 
orange (red) lines represent the swarm model with collection scheme I (II).
The line types indicate the mean number of swarms per grid point
($N_p/N_{\rm grid}$); the total number density
of physical particles is kept the same for all simulations by changing
the number density of particles in each swarm and the number of swarms;
see Runs~4A, 4C, 4D, 4E, 4F and 4G of
Table~\ref{runs_appendix} for simulation details.
Lower panels: similar to the upper panels,
but for $N_p/N_{\rm grid}$=4 and different total numbers of swarms,
as indicated by the line types;
the corresponding $N_{\rm grid}$ is 
$8^3$ (solid curve), $16^3$ (dotted curve),
$32^3$ (dashed curve), $48^3$ (dash-dotted curve) 
and $64^3$ (dash-triple-dotted curve);
see Runs~5A, 5B, 5C, and 5D of
Table~\ref{runs_appendix} for simulation details.
}\label{moments_shima_NpNgrid4}\end{figure}

\section{Statistical convergence of the swarm model}
\label{swarm_statistics}

The purpose of this appendix is to investigate the statistical
convergence with respect to the number of grid cells and swarms.
First we inspect the convergence property of $N_p/N_{\rm grid}$.
The simulations have been performed with $32^3$ grid points and different
average numbers of swarm particles per grid point ($N_p/N_{\rm grid}=2$--$8$).
It can be seen from the upper panels of Figure~\ref{moments_shima_NpNgrid4}
that the swarm simulations with collection scheme~II
almost converge for $N_p/N_{\rm grid}=4$.
The details of these additional runs are summarized in
Table~\ref{runs_appendix}

From the lower panels of Figure~\ref{moments_shima_NpNgrid4}
it can be seen that for 
simulations with $N_p/N_{\rm grid}=4$, the results
are more or less converged when
the total number of swarms reaches $128\times10^3$. 
Since all fluid variables are spatially
uniform in these simulations,
the number of 
grid points has no effect on the fluid. The number of swarms can therefore
be changed by increasing the total number of grid points
while maintaining $N_p/N_{\rm grid}=4$
(the value of $n_i$ is approximately the same in all cases;
$n_i\approx10^9$).
However, as reported by \cite{Arabas13}, when the swarm model is
used in an LES simulation, certain macrophysical features of their
simulated could field does not show convergence regarding grid resolution.

\section{MBR dependency for collection}
\label{MBR_dependency_coag}

\begin{figure}[t!]
\begin{center}
\includegraphics[width=\textwidth]{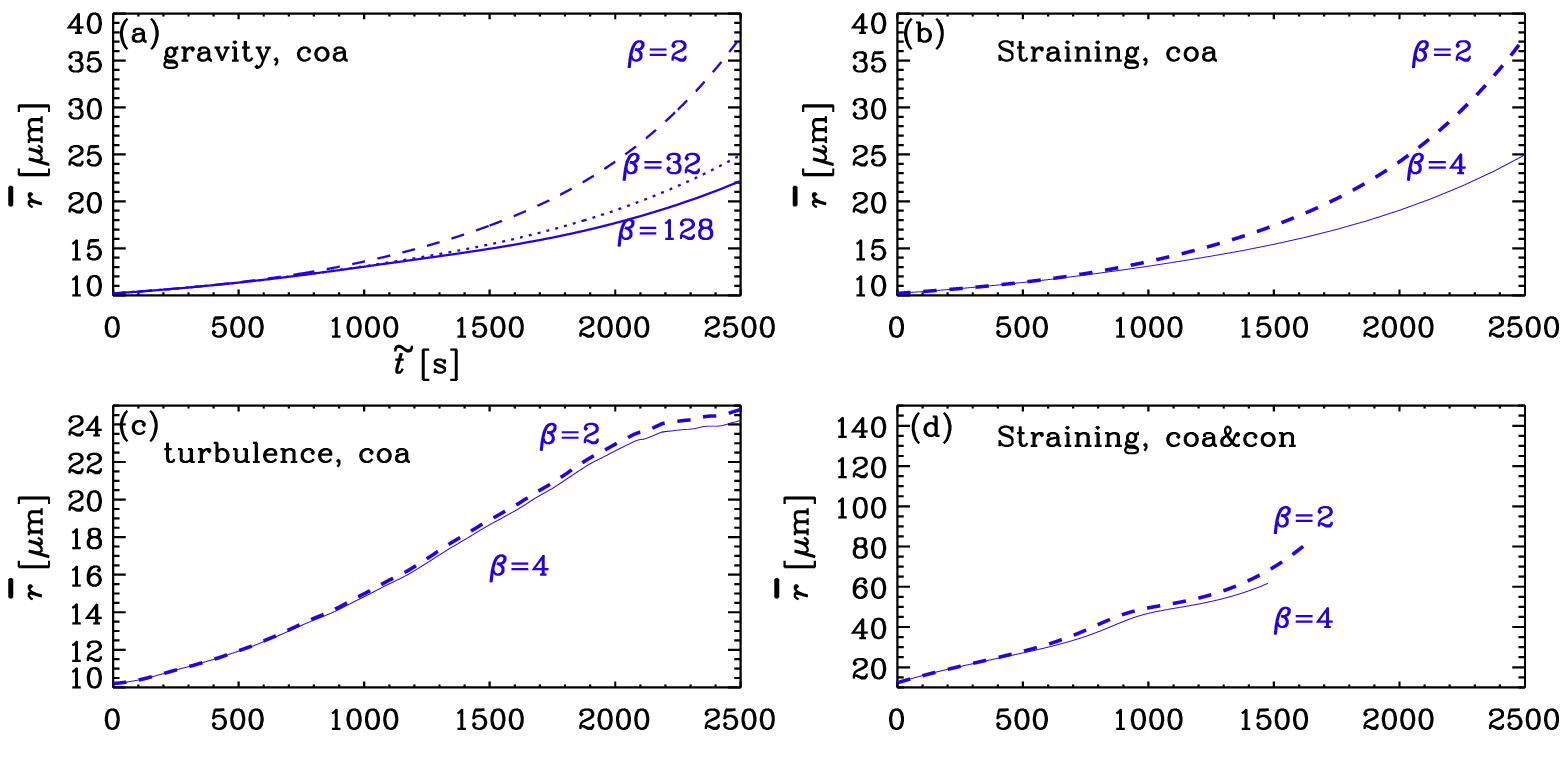}
\end{center}
\caption[]{
MBR dependency for simulations with different flow pattern.
Upper left panel: collection driven by gravity using 
$k_{\max}$=3457 and $\beta$=128 (solid), $k_{\max}$=865 and
$\beta$=32 (dotted), as well as $k_{\max}$=55 and $\beta$=2;
see Runs~3A and 3B of Table~\ref{runs_appendix} and 3C of
Table~\ref{runs} for simulation details.
Upper right panel: collection
driven by straining flow using $k_{\max}$=109 and
$\beta$=4 (dashed line), $k_{\max}$=55 and
$\beta$=2 (solid line); 
see Runs 8A and 8F of Table~\ref{runs_appendix}
for simulation details.
Low left panel: collection driven by turbulence using
$k_{\max}$=109 and $\beta$=4 (dashed line), $k_{\max}$=55
and $\beta$=2 (solid line); 
see Runs 10E of Table~\ref{runs_appendix} and 10A of
Table~\ref{runs} for simulation details.
Low right panel: collection driven by straining flow with
condensation using $k_{\max}$=109 and $\beta$=4 (dashed line),
$k_{\max}$=55 and $\beta$=2 (solid line);
see Runs 8G of Table~\ref{runs_appendix} and 9D of
Table~\ref{runs} for simulation details.
}\label{comp_moments_nocond}
\end{figure}

In Figure~\ref{comp_moments_nocond}, we compare the evolutions of
$\bar{r}$ and $a_{24}$ using different MBR and thus
different values of $\beta$ for pure collection experiment.
We also considered the evolution of $a_3$ and $a_6$,
but it was similar to
that of $a_{24}$ in that the Eulerian solutions for different
MBR agreed well with each other.
For $\bar{r}$ the evolutions are strongly MBR dependent.
Nevertheless, the evolution of $a_{24}$ with 
different MBR are similar over a wide range of 
MBR spanning from $k_{\rm max}=55$--$3457$. 
We also tested the MBR dependency using 
a constant kernel.
In that case, it turns out that the results converge 
only for $k_{\rm max}\ge50$. 

\section{Gamma distribution from higher moments}
\label{app:gamma}

The purpose of this appendix is to show that the characterization
of size spectra in terms of a gamma distribution is not strongly
dependent on whether its parameters are computed based on the
moments $a_1$ and $a_2$, as done here, or based on 
$a_3$ and $a_6$, as done in \cite{Seifert16}.
In the former case, equation~\eqref{shape} was used to obtain
$\mu$ as a function of $a_2/a_1$, while in the latter
$a_6/a_3$ was used to obtain $\mu$.
Once $\mu$ is known. $\lambda$ can also be obtained.
In the following we generalize this approach to arbitrary values
of $\zeta$ for the ratios $a_\zeta/a_{\zeta/2}$.

\begin{figure}[t!]\begin{center}
\includegraphics[width=.7\textwidth]{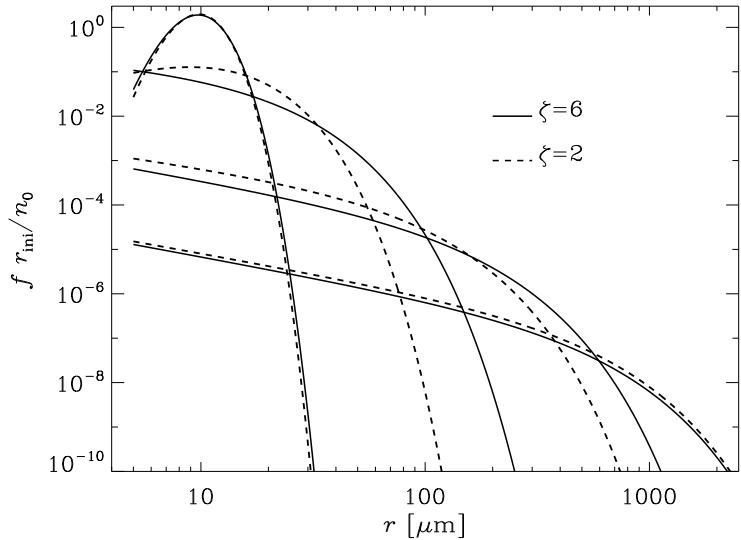}
\end{center}\caption[]{
Similar to Figure~\ref{moments_gamma}, but here we compare
the Eulerian simulation with
gamma distributions with coefficients obtained
from $a_2/a_1$ (solid line) and $a_6/a_3$ (dashed line).
}\label{moments_gamma_multi}
\end{figure}

To compute the coefficients for any pair of moments
$a_\zeta$ to $a_{\zeta/2}$, we first calculate
\begin{equation}
n\,a_\zeta^\zeta=\int r^\zeta f(r)\,{\rm d}r
=\frac{n}{\lambda^\zeta}\,\frac{\Gamma(\mu+\zeta+1)}{\Gamma(\mu+1)}.
\end{equation}
This allows us to derive the following general formula
\begin{equation}
\frac{a_\zeta}{a_{\zeta/2}}
=\left[\frac{\prod_{\zeta'=\zeta/2+1}^{\zeta}(\mu+\zeta')}
{\prod_{\zeta'=\zeta/2+1}^{\zeta}(\mu+\zeta')}\right]^{1/\zeta}.
\end{equation}
The expression on the right-hand side of this equation is a monotonic
function of $\mu$ and can easily be solved numerically.
Once we know $\mu$, we compute $\lambda$ via
\begin{equation}
\lambda=\frac{1}{a_\zeta}\,\left(\frac{\Gamma(\mu+\zeta+1)}{\Gamma(\mu+1)}\right)^{1/\zeta}
\end{equation}
The resulting pairs of coefficients $(\mu_\zeta,\lambda_\zeta)$
are given in Table~\ref{Tfit} for Run~3C for different times
and several values of $\zeta$.
The moments used here are given in Table~\ref{TamomText}.
The result for $a_6/a_3$ is shown in Figure~\ref{moments_gamma_multi},
where we also compare with the result for $a_2/a_1$.
The agreement between the two is surprisingly good.

\begin{table}[t!]\caption{
Results for $(\mu_\zeta,\lambda_\zeta)$ for different times.
}\vspace{12pt}\centerline{\begin{tabular}{crrrrr}
$t$ & $2\quad\quad\quad$ & $4\quad\quad\quad$ & $6\quad\quad\quad$ & $12\quad\quad\quad$ & $24\quad\quad\quad$ \\
\hline
0 & $(23.61,\,2.412)$&$(22.50,\,2.306)$&$(21.57,\,2.218)$&$(18.74,\,1.967)$&$(13.47,\,1.547)$\\
1 & $( 2.16,\,0.242)$&$( 0.20,\,0.108)$&$(-0.33,\,0.079)$&$( 0.03,\,0.093)$&$( 3.25,\,0.159)$\\
2 & $(-0.69,\,0.017)$&$(-0.91,\,0.009)$&$(-0.88,\,0.010)$&$(-0.49,\,0.014)$&$( 1.91,\,0.023)$\\
3 & $(-0.92,\,0.003)$&$(-0.94,\,0.002)$&$(-0.90,\,0.003)$&$(-0.33,\,0.005)$&$( 4.39,\,0.010)$\\
\label{Tfit}\end{tabular}}\end{table}

\section{The ``bump'' in the evolution of the mean particle radius}
\label{app:bump}

For the following discussion, it is convenient to introduce the unscaled moments
\begin{equation}
M_\zeta=\sum f(r)\, r^\zeta.
\label{nzeta}
\end{equation}
so that $a_\zeta=(M_\zeta/M_0)^{1/\zeta}$ and $\overline{r}=a_1$, as before.
Let us now assume a situation with pure collection
such that the total volume of water in the droplets is conserved. This 
implies that $M_3$ is constant, while $M_0$ and $M_1$ will always
decrease with time.
However, the relative rates at which $M_0$ and $M_1$ decrease can change.
Indeed, a bump in $\overline{r}$ is observed if $M_1$ switches from decreasing more
slowly with time than $M_0$ to decreasing faster than $M_0$.
An example of such a situation will be presented in the following.

\begin{figure}[t!]\begin{center}
\includegraphics[width=.8\textwidth]{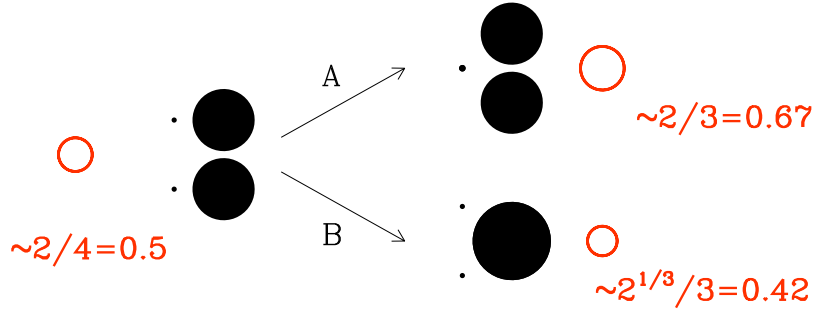}
\end{center}\caption[]{Sketch illustrating the growth
of $\overline{r}$ when two small particles collide ({\sf A}) and the
decrease of $\overline{r}$ when two large particles collide ({\sf B}).
Filled black symbols denote actual particle sizes and
open red symbols and red text refer to $\overline{r}$.
}\label{sketch}\end{figure}

For a flow with two small and two large particles, with radii $r_{\rm S}$ and 
$r_{\rm L}$, respectively, the size distribution is given by
$f(r)=2\delta_{r\,r_{\rm S}}+2\delta_{r\,r_{\rm L}}$,
where $\delta_{ij}$ denotes the Kronecker delta
($\delta_{ij}=1$ for $i=j$ and $0$ otherwise).
From equation~\ref{nzeta} it can then be found that the initial number of
particles and sum of particle radii is given by $M_0(0)=4$ and 
$M_1(0)=2r_{\rm S}+2r_{\rm L}$, respectively.
This yields a mean initial particle radius of 
$\overline{r}(0)=M_1(0)/M_0(0)$.
In the following, we assume that $r_{\rm S}\ll r_{\rm L}$,
so that $\overline{r}(0)\approx2r_{\rm L}/4=0.5r_{\rm L}$.

When two particles of radius $r_0$ collide, their combined mass is unchanged,
so $2r_0^3=r^3$, i.e., the target radius becomes $r=2^{1/3}r_0$ \citep{LV11}.
Let us now consider two different collection scenarios; cf.\ Figure~\ref{sketch}.
In scenario~{\sf A}, two smaller particles collide such that 
$M_0({\sf A})=3$ and $M_1({\sf A})=2^{1/3}r_{\rm S}+2r_{\rm L}$,
while in scenario~{\sf B} two larger particles collide such that
$M_0({\sf B})=3$ and $M_1({\sf B})=2r_{\rm S}+2^{1/3}r_{\rm L}$.
Since $r_{\rm L}\gg r_{\rm S}$, we find for $\overline{r}$ in both scenarios
\begin{equation}
\overline{r}({\sf A})
=(2^{1/3}r_{\rm S}+2r_{\rm L})/3
\approx 2r_{\rm L}/3\approx0.67r_{\rm L} > \overline{r}(0),
\end{equation}
\begin{equation}
\overline{r}({\sf B})
=(2r_{\rm S}+2^{1/3}r_{\rm L})/3
\approx 2^{1/3}r_{\rm L}/3\approx0.42r_{\rm L}<\overline{r}(0).
\end{equation}
This means that for scenario {\sf A} the mean particle
radius is increasing, while for scenario {\sf B} it is decreasing.
After the time when the bump appears in the time
evolution of the mean particle radius (see
Figure~\ref{pcomp_L72condens22coag}),
it is primarily
the heavier particles that continue collecting.

\section{Self-collection in the Eulerian model}
\label{SelfCollision}

\begin{figure}[t!]\begin{center}
\includegraphics[width=\textwidth]{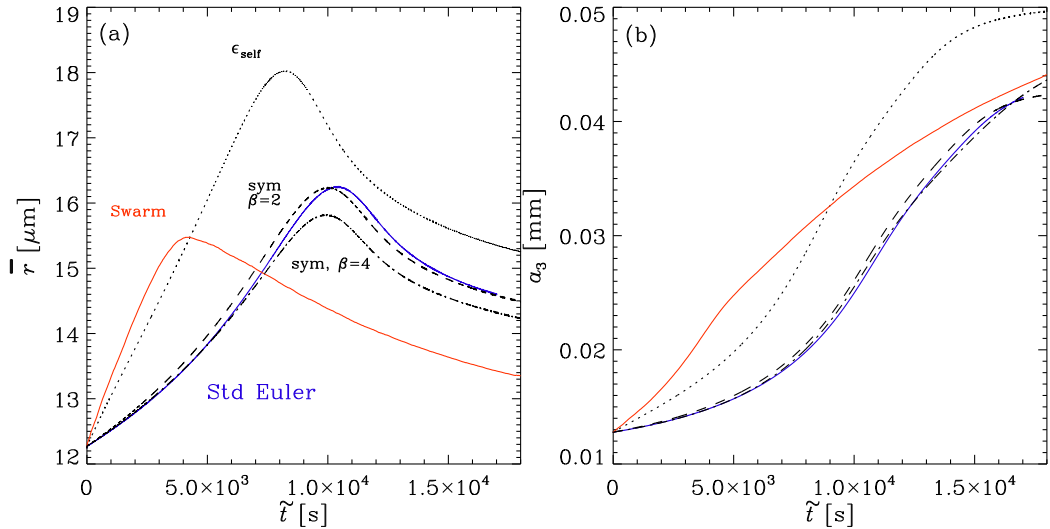}
\end{center}\caption[]{
Similar to the Eulerian model in Figure~\ref{pcomp_amean_L256condens53coag_forceStrain},
but with self-collection invoked.
Comparison of purely collectional growth 
for Eulerian models with average self-collection ($\epsilon_{\rm self}$=0.01) and
symmetric self-collection (dashed for $\beta$=2 with $k_{\max}$=53 and dash-dotted for
$\beta$=4 with $k_{\max}$=109), as well as the swarm model in a straining flow.
Here, $u_{\rm rms}=0.7\,{\rm m}\,{\rm s}^{-1}$, $\tau_{\rm cor}=1.4\,{\rm s}$, while
$\tau_{\rm coll}\approx100\,{\rm s}$. 
The side length of the 2-D squared domain is $L$=$2\pi\,{\rm m}$.
The parameters of the Eulerian model are
$n_0$=$10^{10}\,{\rm m}^{-3}$, $r_1=4\,\mu{\rm m}$, and $r_{\rm ini}=12\,\mu{\rm m}$.  
Those for the swarm model are $N_p=50000$ and NGP mapping is employed.
See Runs~8A, 8B, 8C, 8D, and 8E of
Table~\ref{runs} for simulation details.
}\label{pcomp_Euler_swarm_cond0_coag_kin_self_xy}\end{figure}

We recall that there are no self-collections in the usual Eulerian scheme.
The potential importance of this can be assessed by comparing
with calculations in which self-collection is included either
via methods (i) or (ii); see the end of section~\ref{eul_app} for their definitions.
It turns out that by taking self-collection into account, method (i)
causes only a weak speed-up in the increase of $\overline{r}$;
see Figure~\ref{pcomp_Euler_swarm_cond0_coag_kin_self_xy}.
With method (ii), on the other hand,
we find a strong enhancement of the growth.
However, although method (ii) consists of an artificial manipulation of
the diagonal terms of $K_{ij}$, it does not prove that self-collection
is important, because similar manipulations of the off-diagonal terms
of $K_{ij}$ can have the same effect.
In any case, this unphysical approach does not provide a proper solution
to the convergence problem. 
We mention in passing that for this straining flow the
effect of the ${\cal M}_k$ term in equations~\eqref{dvdt}
and (\ref{Mkterm}) has no effect within plot accuracy.

\end{appendices}


\end{document}